\documentclass[fleqn,usenatbib]{mnras}

\usepackage{newtxtext,newtxmath}
\usepackage[T1]{fontenc}
\usepackage{ae,aecompl}
\usepackage{graphicx}	
\usepackage{amsmath}
\usepackage{adjustbox}
\usepackage{hyperref}
\usepackage{url}

\title[The SMEI view of CP2/4 stars]{Magnetic chemically peculiar stars investigated by the Solar Mass Ejection Imager}

\author[E.~Paunzen et al.]{E.~Paunzen,$^{1}$\thanks{E-mail: epaunzen@physics.muni.cz}
   J.~Sup{\'i}kov{\'a},$^{1,2}$
   K.~Bernhard,$^{3,4}$
   S.~H{\"u}mmerich,$^{3,4}$
   M. Pri{\v s}egen$^{1}$\
	\\
$^{1}$Department of Theoretical Physics and Astrophysics, Masaryk University, Kotl{\'a}\v{r}sk{\'a} 2, CZ\,611\,37 Brno, Czech Republic\\
$^{2}$Faculty of Informatics, Masaryk University, Brno, Czech Republic\\
$^{3}$Bundesdeutsche Arbeitsgemeinschaft f{\"u}r Ver{\"a}nderliche Sterne e.V. (BAV), Berlin, Germany\\
$^{4}$American Association of Variable Star Observers (AAVSO), Cambridge, USA\\
}
	
\date{Accepted XXX. Received YYY; in original form ZZZ}
\pubyear{2021}

\begin{document}
\label{firstpage}
\pagerange{\pageref{firstpage}--\pageref{lastpage}}
\maketitle

\begin{abstract}
Since the discovery of the spectral peculiarities of their prototype $\alpha^2$ Canum Venaticorum in 1897, the so-called ACV variables, which are comprised of several groups of chemically peculiar stars of the upper main sequence, have been the target of numerous photometric and spectroscopic studies. Especially for the brighter ACV variables, continuous observations over about a century are available, which are important to study long-term effects such as period changes or magnetic cycles in these objects. The present work presents an analysis of 165 Ap/CP2 and He-weak/CP4 stars using light curves obtained by the Solar Mass Ejection Imager (SMEI) between the years 2003 and 2011. These data fill an important gap in observations for bright ACV variables between the {\it Hipparcos} and TESS satellite missions. Using specifically tailored data treatment and period search approaches, we find variability in the accuracy limit of the employed data in 84 objects. The derived periods are in excellent agreement with the literature; for one star, the here presented solution represents the first published period. We discuss the apparently constant stars and the corresponding level of non-variability. From an investigation of our target star sample in the Hertzsprung-Russell diagram, we deduce ages between 100 Myr and 1 Gyr for the majority of our sample stars. Our results support that the variable CP2/4 stars are in a more advanced evolutionary state and that He and Si peculiarities, preferentially found in the hotter, and thus more massive, CP stars, produce larger spots or spots of higher contrast.
\end{abstract}


\section{Introduction} \label{introduction}

Chemically peculiar (CP) stars form a significant fraction (up to 15\%) of upper main-sequence stars and are mostly found between spectral types early B to early F. The defining characteristic of CP stars is the presence of spectral peculiarities, which indicate unusual elemental abundance patterns. Several groups of CP stars have been defined, such as the metallic-line or Am (CP1) stars, the magnetic Bp/Ap (CP2) stars, the Mercury-Manganese (HgMn/CP3) stars, and the He-weak (CP4) stars \citep{1974ARA&A..12..257P,2018MNRAS.480.2953G,2019MNRAS.487.5922G}.

The CP2 and CP4 stars are distinguished from the CP1 and CP3 stars by the presence of globally organized magnetic fields with strengths of about 300\,G to several tens of kiloGauss \citep{2008AstBu..63..139R}. \citet{1950MNRAS.110..395S} introduced the Oblique Rotator model of magnetic stars, which assumes non-coincidence of magnetic and rotational axes and is able to reproduce the observed variability and the reversals of the magnetic field strength. Due to chemical abundance concentrations at the magnetic poles, CP2/4 stars also show spectral and photometric variability, as well as radial velocity variations of the appearing and receding patches on the stellar surface, which are also easily understood in terms of the Oblique Rotator model. Variable CP2/4 stars are traditionally referred to as $\alpha^2$ Canum Venaticorum (ACV) variables \citep{2017ARep...61...80S} and are characterised by light curves that remain stable over decades or more.

There is a long tradition of photographic and photoelectric investigations of bright ACV variables in the literature \citep{1962ApJ...136...35A,1970ApJ...161..685P,1981A&AS...46..151H,1993ASPC...44..644A,2018PASP..130d4202D}. The present work concentrates on the analysis of bona-fide CP2 and CP4 stars using light curves from the Solar Mass Ejection Imager \citep[SMEI,][]{2003SoPh..217..319E}, which continuously observed bright stars ($V$\,$<$\,7\,mag) from 2003 to 2011.

For the analysis of the long-term behaviour of chemical spots on the surface of CP2/4 stars, it is essential to have photometric observations with a long time baseline. Effects such as magnetic cycles and long-term variations of the rotational period are rarely investigated, and no data are currently available for a statistically sound sample of these stars. For example, precise observations with long time baselines are needed for the analysis of rotational period changes in ACV variables, and only very few of these stars are currently known which underwent measurable changes \citep{2016CoSka..46...95M}.

If an analysis of the long-term behaviour of ACV variables is to include data from photographic plates, such as the data provided the Digital Access to a Sky Century@Harvard (DASCH) project \citep{2013PASP..125..857T}, we are mainly limited to bright stars. For these objects, many photoelectric studies, beginning in the 1950s, are available. However, since the beginning of CCD photometry, observations of bright stars have become scarce in the literature.

Before this background, the photometric observations of the SMEI satellite fill an important gap in time between the {\it Hipparcos} mission \citep{1994SoPh..152...91E} and the observations of the Transiting Exoplanet Survey Satellite \citep[TESS,][]{2015JATIS...1a4003R}, which started in April 2018 and is observing stars as bright as 4th magnitude. The bright objects targeted by these missions have not been observed by the All Sky Automated Survey \citep[ASAS-3,][]{2002AcA....52..397P}, the Wide Angle Search for Planets \citep[WASP,][]{2006PASP..118.1407P}, and similar surveys because of saturation issues. 

In general, TESS data are superior in precision to the SMEI data. However, for about 25\,\% of our sample stars, no TESS photometry is available yet. Furthermore, the time baseline of the TESS data is still short and, in many cases, covers only some rotational cycles. The time baseline of the SMEI data, on the other hand, is much longer and therefore better suited to derive rotational periods with a precision sufficient to our goals. In addition to that, several bright stars show saturation issues in TESS data, which are not observed in the SMEI data sets. The analysis of SMEI data, therefore, is important for the study of bright ACV variables, for which ground-based follow-up spectroscopic and spectropolarimetric observations are comparatively easy to achieve.


In total, we analysed the light curves of 165 stars, 84 of which show variability in the accuracy limits of the SMEI data. Except for one star, all variables have literature periods, so the data can be employed for the study of their long-term behaviour. We furthermore discuss the noise limits of the amplitude spectra of the non-variable CP2/4 stars. We emphasise that the recent papers discussing the characteristics of ACV variables \citep{2015A&A...581A.138B,2016AJ....152..104H,2020MNRAS.493.3293B} do not derive any statistics on non-variable CP2/4 stars which should also have spots on their surfaces and thus should show variability. The boundary conditions of when ACV variability occurs are, in general, poorly studied. 

In addition, we discuss the location of our target star sample within the classical Hertzsprung-Russell diagram and compare the available parallaxes and luminosities of the {\it Hipparcos} and {\it Gaia} missions.

\begin{figure}
\begin{center}
\includegraphics[width=0.49\textwidth]{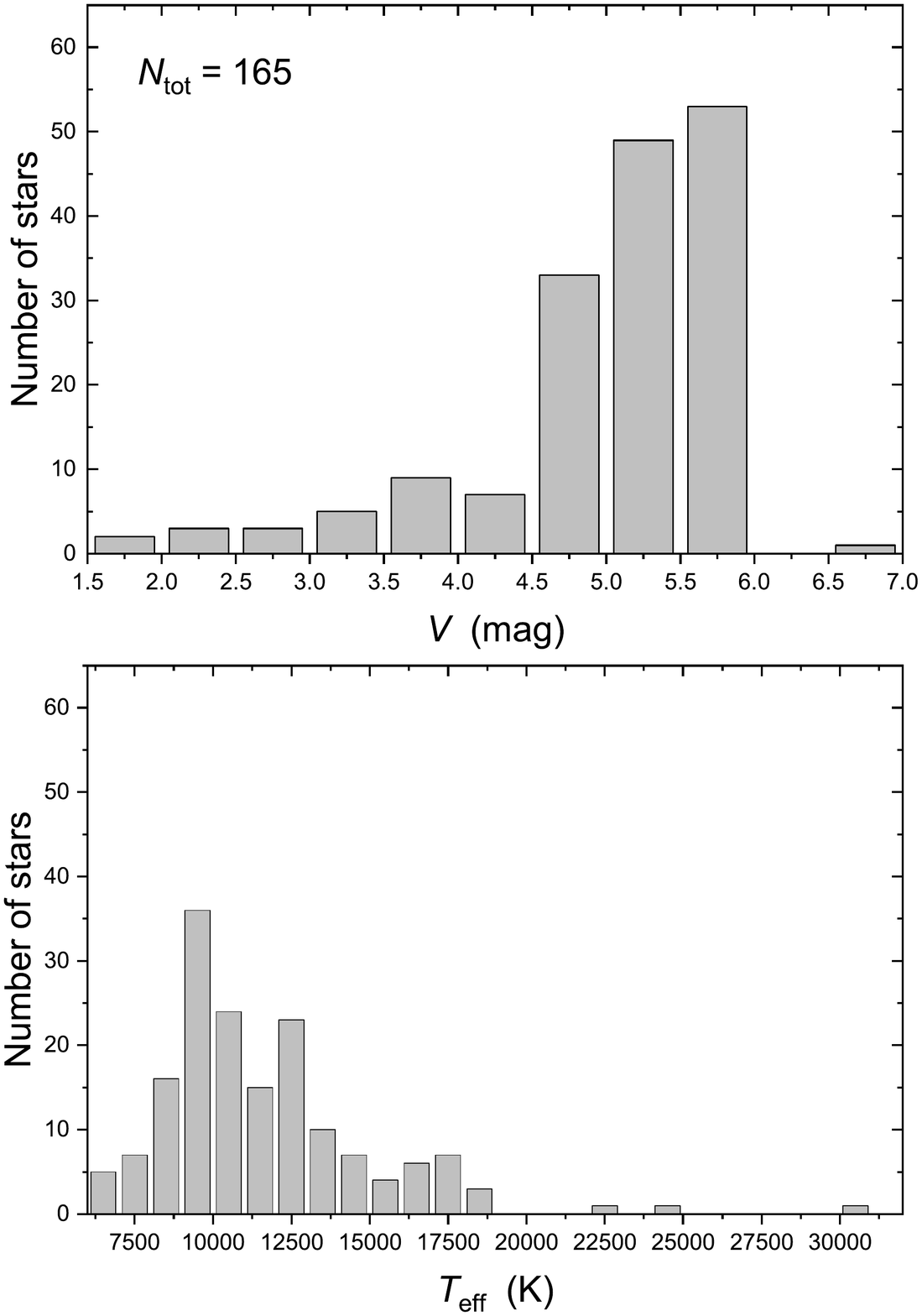}
\caption{Histograms of the $V$ magnitudes (upper panel) and effective temperatures (lower panel) of our target star sample.}
	\label{histograms}
\end{center}
\end{figure}

\section{Data sources and target selection} \label{data_selection}

We searched for all probable CP2 and CP4 stars in the most up-to-date collection, the most recent version of the General Catalogue of CP Stars published by \citet{RM09}. Because SMEI observed only bright, and hence for the most part well studied, stars (Fig. \ref{histograms}), the classifications in this catalogue should be reliable. However, we also included some borderline cases for which the spectral classification is not clear. Here we recall the remarks of \citet{RM09} on the classification of CP stars: ''/'' -- a star improperly considered Ap or Am; ''?'' -- CP star of doubtful nature; and ''*'' -- well-known Ap or Am star. Although these classifications have to be viewed with caution, we excluded all stars denoted with a ''/''. In total, 165 stars were selected as our final sample. Essential data for these objects are given in Table \ref{table_master1}. 

The Solar Mass Ejection Imager \citep[SMEI,][]{2003SoPh..217..319E} was launched into an Earth-terminator, Sun-synchronous, 840 km polar orbit as a secondary payload on board the Coriolis spacecraft in January 2003 and was terminated in September 2011. Its main purpose was to monitor and predict space weather in the inner solar system.

SMEI comprised three wide-field cameras, which were aligned such that the total field of view is a 180 deg and about 3 deg wide arc, yielding a near-complete image of the sky after about every 101.5 min orbit. The photometric passband ranges from 450 to 950\,nm. Several papers \citep[for example,][]{2019MNRAS.485.3544W,2019A&A...627A..28Z} successfully employed the long time-baseline SMEI photometry for the study of variable stars.

A detailed description of the data analysis pipeline used to extract light curves from the raw data is provided by \citet{2007SPIE.6689E..0CH}. Stellar time series can be extracted from the SMEI website\footnote{\url{http://smei.ucsd.edu/}}.

\section{Time Series Analysis} \label{tsa}

The time series analysis described in this section was performed with the program packages PERANSO \citep{2016AN....337..239P} and VARTOOLS \citep{2016A&C....17....1H}. For the identification of significant periods, the Generalized Lomb-Scargle algorithm \citep[L-S, ][]{2009A&A...496..577Z} and the Phase-Dispersion Method \citep[PDM,][]{1978ApJ...224..953S} were applied. Both methods complement each other well, especially in the case of the SMEI data sets which are not free of various instrumental effects
(Sect. \ref{instrumental_effects}).

The PDM algorithm is a classical string-length method. First of all, the data are folded on a series of trial periods. For each of these, the original data are assigned phases which are then re-ordered in ascending sequence. The reordered data are examined by inspection across the full phase interval between zero and one. For each trial period, the full phase interval is divided into a number of bins. Then, the sum of the lengths of line segments joining successive points (the string-length) is calculated. The variance of each of these bins is derived, giving a measure of the scatter around the mean light curve, defined by the means of the data in each sample. The PDM statistic is then calculated by dividing the overall variance of all the samples by the variance of the original (unbinned) data set. This process is repeated for each consecutive trial period. Minima in the plot of the string-length versus the trial period can be considered as corresponding to the underlying period.

The L-S algorithm is a variation of the Discrete Fourier Transform (DFT) which converts a finite list of equally spaced samples of a function (here the light curve) into a list of coefficients of a linear combination of sinusoidal and cosinusoidal functions. The data are transformed from the time to the frequency domain (Lomb-Scargle periodogram), invariant to time shifts. From a statistical point of view, the resulting periodogram is related to the $\chi^2$ for a least-squares fit of a single sinusoid to the data which can treat heteroscedastic measurement uncertainties. The underlying model is non-linear in frequency and the basis functions at different frequencies are not orthogonal.

For interpreting the significance of the detected periods, we used the False-Alarm probability (FAP), which denotes the probability that at least one out of $N$ independent power values in a prescribed search band of a power (or amplitude) spectrum computed from a white-noise time series is expected to be as large as or larger than a given value. Its correct interpretation has been often discussed in the literature \citep{1986ApJ...302..757H,1999ApJ...526..890C}. Here, we consider a period significant if the $\log$\,FAP value is larger than 1000. Furthermore, we require that this period is unambiguously detected with both the L-S and PDM methods. These are very strict limits that we feel justified to apply due to the characteristics of the SMEI light curves.

\begin{figure}
\begin{center}
\includegraphics[trim = 0mm 0mm 10mm 125mm, clip, width=0.49\textwidth]{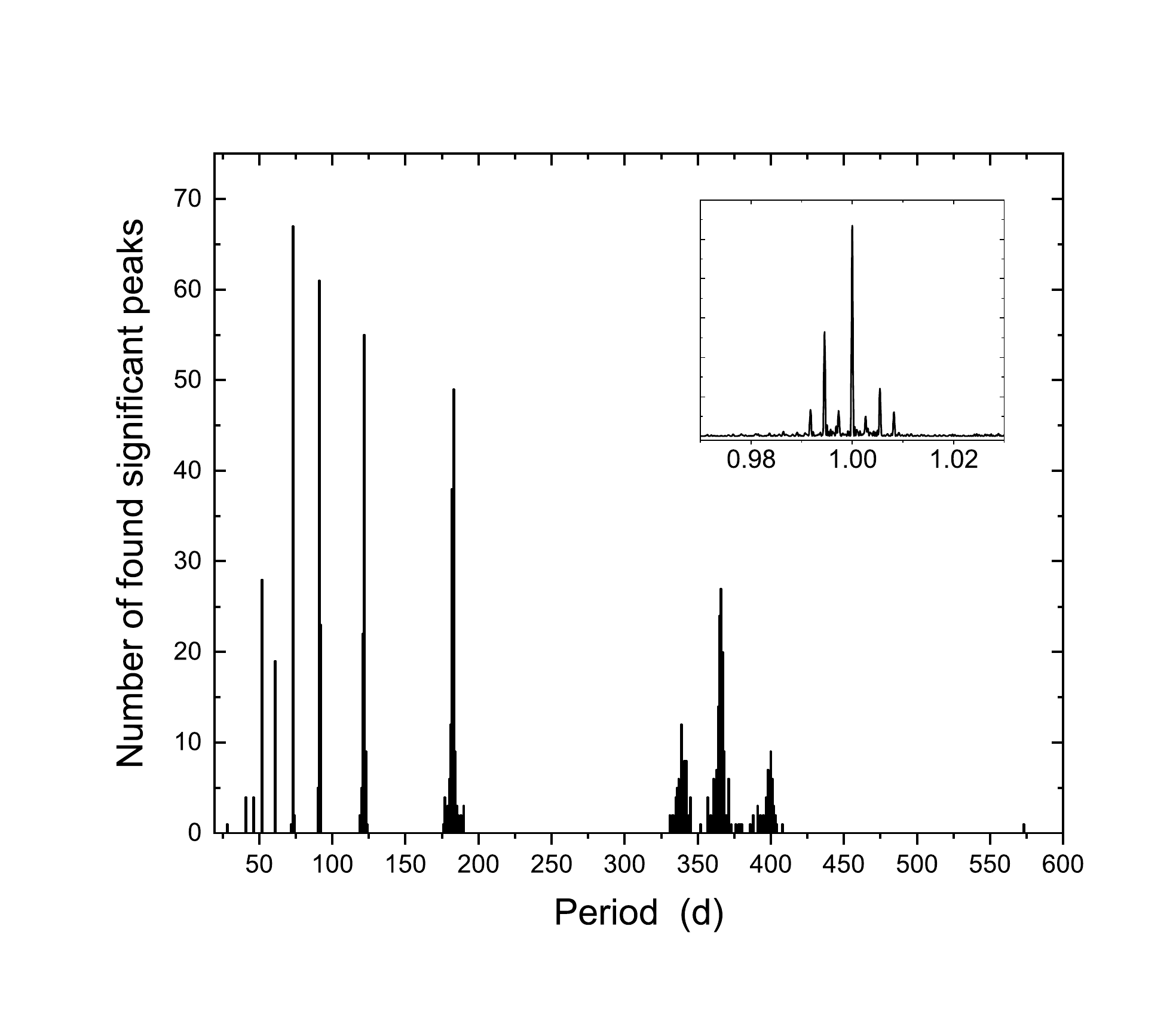}
\caption{Histogram of detected instrumental periods in the range from 20 to 1000 days. Only periods detected more than two times are shown (there are none in the range from 600 to 1000 days).  The panel in the right upper corner shows a typical L-S periodogram around the period of one day.}
	\label{instrumental_periods}
\end{center}
\end{figure}

\begin{figure}
\begin{center}
\includegraphics[width=0.49\textwidth]{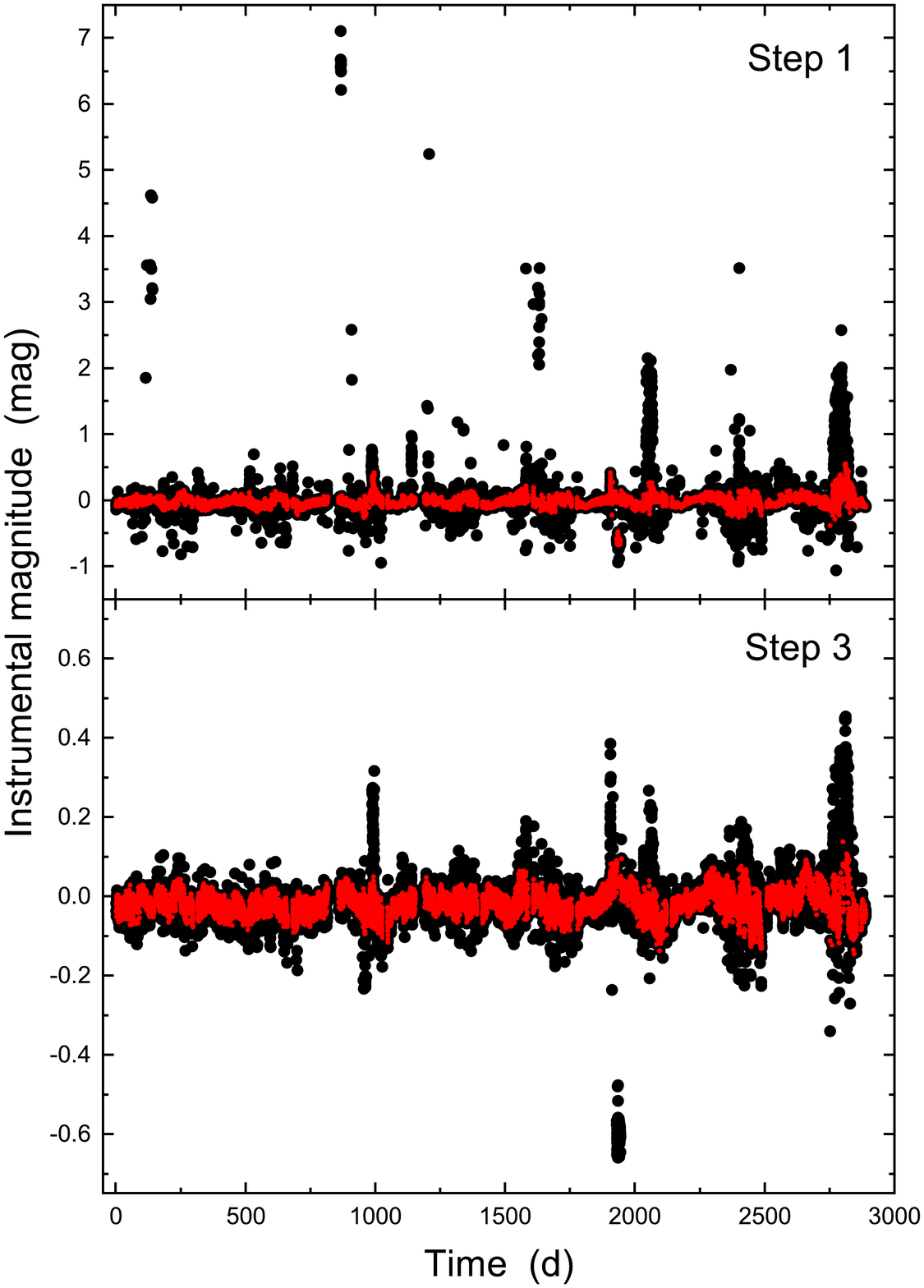}
\caption{Example of the reduction steps one and three as described in Section \ref{instrumental_effects}. The light curve is that of the star HD 128898 ($\alpha$ Cir). Red dots
denote magnitudes after the application of the corresponding reduction step.}
	\label{reduction_steps}
\end{center}
\end{figure}

\subsection{Instrumental effects and data cleaning} \label{instrumental_effects}

SMEI light curves are strongly influenced by the Sun, Moon and daily cycles. The period range of ACV variables spans several magnitudes, extending from about half a day \citep{2017MNRAS.466.1399H} to hundreds or even thousands of years \citep{mathys17}. However, most ACV variables show periods of days \citep{2017MNRAS.468.2745N}. Because of this, the SMEI orbital period of 101.5 min, which introduces artificial signals in the high frequency range, should not influence our results significantly.

To further investigate this matter, all 165 light curves were searched to identify the most important instrumentally-induced periods. The L-S algorithm was used to search for signals in the range from 20 to 1000 days, and the data were consecutively pre-whitened with the most significant periods, after which the L-S algorithm was applied again. This procedure was carried out five times.

In Fig. \ref{instrumental_periods}, we present the histogram (binned to one day intervals) of the derived instrumental periods. Only periods detected more than two times are shown; none were found in the range from 600 to 1000 days. In 111 out of the 165 light curves, the period around one year is the most significant one, followed by the half year period for an additional 47 light curves. Overall, we found significant peaks at around 52, 61, 73, 91, 122, 183, and 366 days. We also note the strong ''side lobes'' of the one year period around 339 and 400 days. In addition, a one day period with a typical pattern is clearly visible in the periodograms (Fig. \ref{instrumental_periods}, upper right corner). 

All cleaning and time series analysis steps are demonstrated using the light curve of HD 128898 ($\alpha$ Cir), which is a rapidly oscillating Ap (roAp) star that also shows low-amplitude ACV variations \citep{2009MNRAS.396.1189B}. Together with photometric data from the BRITE Constellation, TESS, and WIRE satellites, its SMEI light curve was used in the recent paper of \citet{2020A&A...642A..64W} to study the spot structure of the star. Unfortunately, in their paper, the authors do not show amplitude spectra based on SMEI data (see Figs. 5 and 6 therein). The following automatic procedure was applied to all light curves.

\begin{itemize}
    \item Step 1: cleaning of the raw light curve using a basic $\sigma$-clipping algorithm
    \item Step 2: consecutive prewhitening of the five periods with the highest FAP in the range from 50 to 1000 days and the one day period
    \item Step 3: cleaning of the residual light curve using a basic $\sigma$-clipping algorithm
    \item Step 4: search for significant periods
\end{itemize}

Fig. \ref{reduction_steps} illustrates the first and third steps graphically. It also highlights the strong instrumental trends present in the raw light curve.

\begin{figure}
\begin{center}
\includegraphics[width=0.49\textwidth]{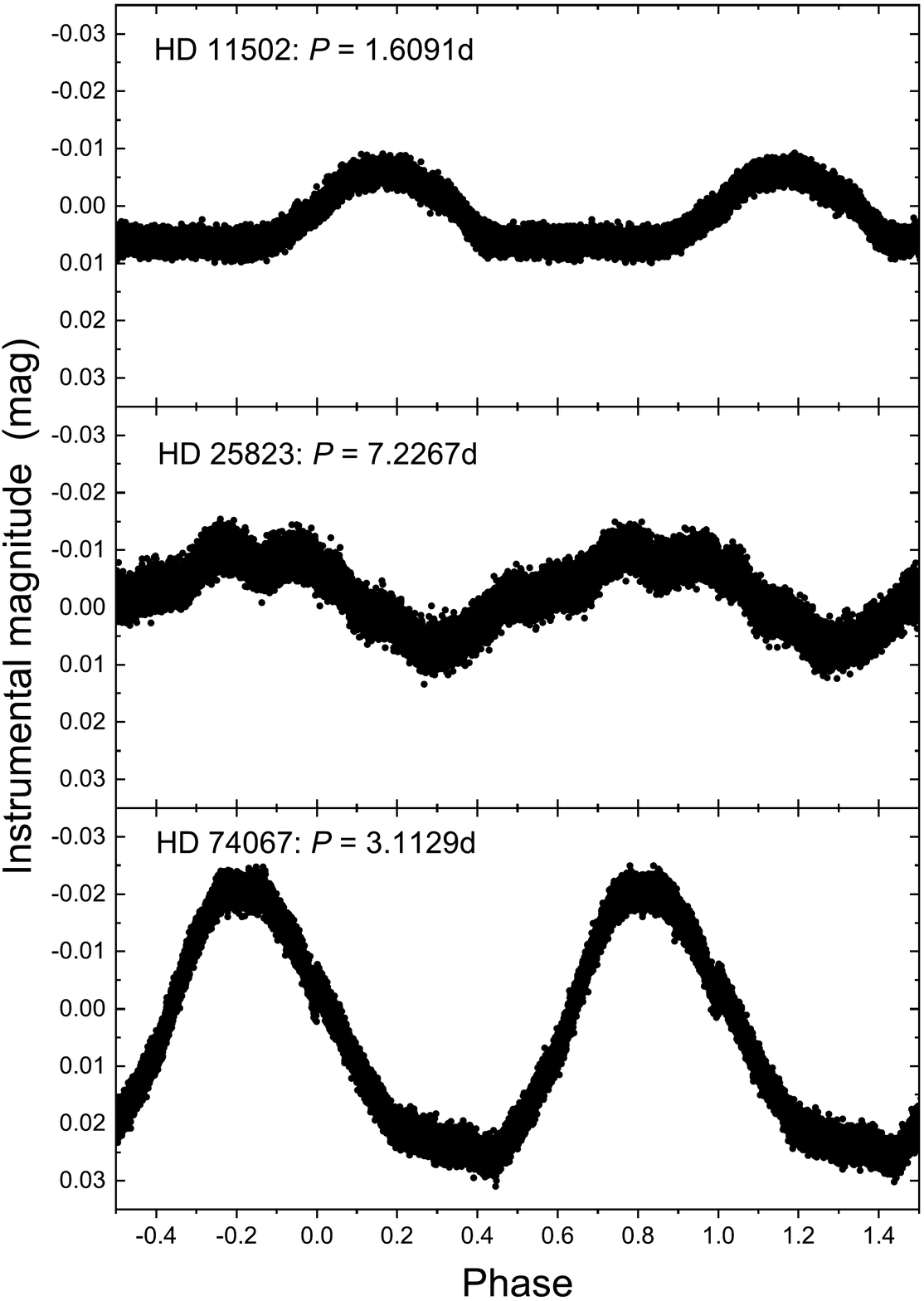}
\caption{Phased SMEI light curves of three sample stars with widely different periods and light curve characteristics.}
	\label{phased_light_curves}
\end{center}
\end{figure}

\begin{figure}
\begin{center}
\includegraphics[trim = 0mm 0mm 10mm 125mm, clip, width=0.49\textwidth]{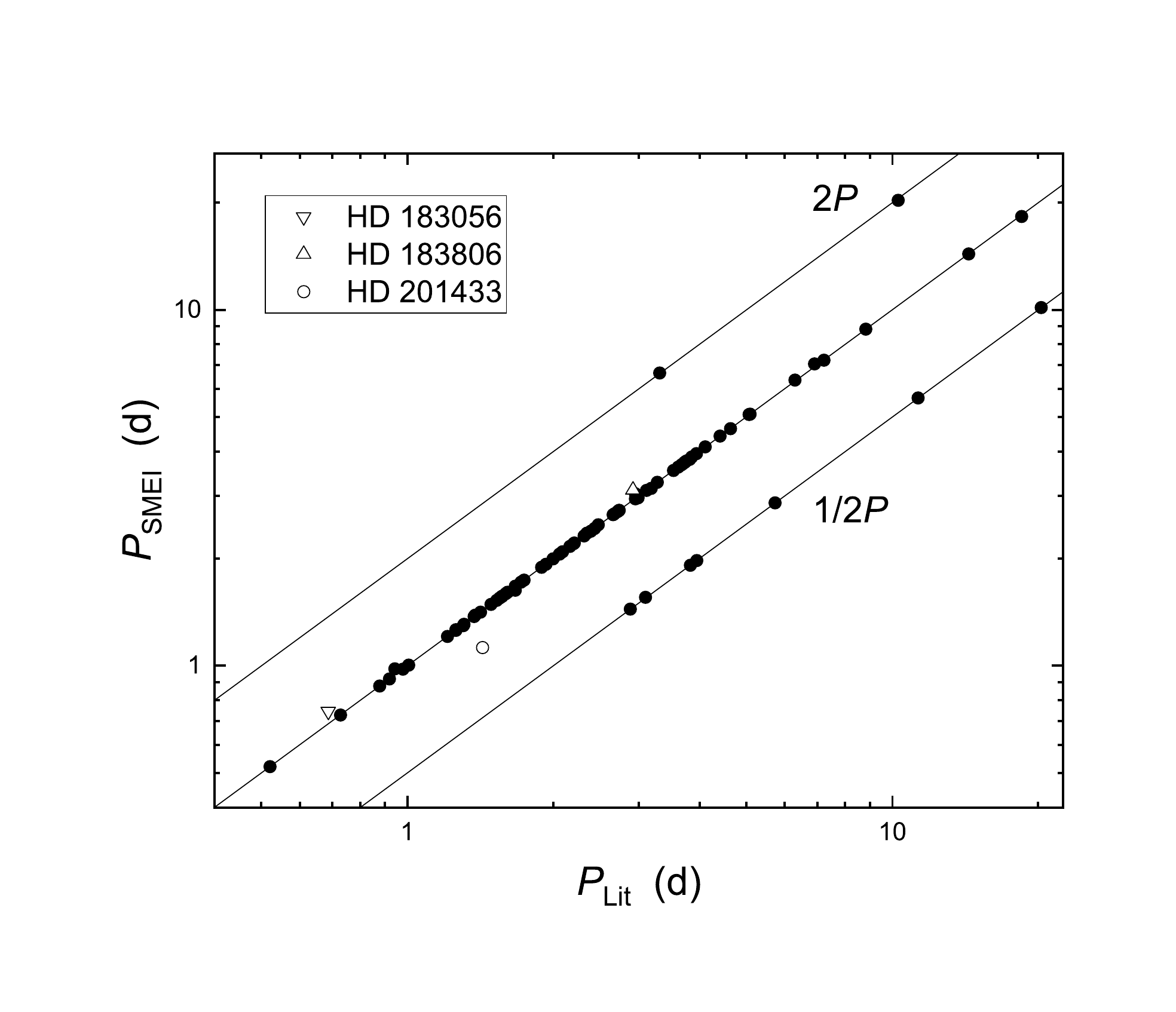}
\caption{Comparison of the period values derived from SMEI data to the literature periods for our sample stars. Also indicated are the unity line and the lines correspondings to the half- and double-periods. Open symbols denote the three outliers (HD 183056, HD 183806, and HD 201433), which are discussed in the text.}
	\label{comparison_known_variables}
\end{center}
\end{figure}

\begin{table*}
\begin{center}
\caption{Significant periods of the 84 stars of our target star sample that show variability in the accuracy limit of the SMEI data, as based on the time-series analysis described in Section \ref{tsa}. Literature periods were gleaned from the International Variable Star Index \citep[VSX,][]{2006SASS...25...47W} and the compilations of \citet{2017MNRAS.468.2745N} and \citet{2019A&A...622A.199J}. The typical period error amounts to 0.0001\,d.}
\label{found_periods}
\begin{tabular}{ccc|ccc|ccc|ccc}
  \hline
{\bf HD} & $P_\mathrm{SMEI}$ & $P_\mathrm{Lit}$ & {\bf HD} & $P_\mathrm{SMEI}$ & $P_\mathrm{Lit}$ & {\bf HD} & $P_\mathrm{SMEI}$ & $P_\mathrm{Lit}$ & {\bf HD} & $P_\mathrm{SMEI}$ & $P_\mathrm{Lit}$ \\
& (d) & (d) &  & (d) & (d) &  & (d) & (d) &  & (d) & (d) \\
\hline
{\bf 3980} & 1.9758 & 3.9516 & {\bf 36485} & 2.8662 & 5.7325 & {\bf 96616} & 2.4292 & 2.4330 & {\bf 148112} & 3.0437 & 2.9510 \\
{\bf 10221} & 3.1538 & 3.1848 & {\bf 39317} & 2.6571 & 2.6569 & {\bf 103192} & 2.3567 & 2.3440 & {\bf 148898} & 2.9642 & 2.9900 \\
{\bf 11502} & 1.6091 & 1.6092 & {\bf 40312} & 3.6186 & 3.6187 & {\bf 108662} & 5.0761 & 5.0633 & {\bf 150549} & 3.7597 & 3.7600 \\
{\bf 12767} & 1.8923 & 1.8923 & {\bf 40394} & 14.3661 & 14.3680 & {\bf 109026} & 2.7280 & 2.7293 & {\bf 151525} & 4.1163 & 4.1160 \\
{\bf 14392} & 1.3086 & 1.3080 & {\bf 47144} & 2.2105 & 2.2100 & {\bf 112185} & 5.0889 & 5.0886 & {\bf 152107} & 3.8575 & 3.8567 \\
{\bf 15089} & 1.7403 & 1.7405 & {\bf 47152} & 2.7112 & 2.7112 & {\bf 118022} & 3.7225 & 3.7220 & {\bf 152564} & 2.1635 & 2.1637 \\
{\bf 18296} & 1.4421 & 2.8842 & {\bf 54118} & 3.2752 & 3.2753 & {\bf 120198} & 1.3858 & 1.3800 & {\bf 166596} & 1.6299 & 1.6700 \\
{\bf 19832} & 0.7279 & 0.7279 & {\bf 56022} & 0.9184 & 0.9183 & {\bf 124224} & 0.5207 & 0.5207 & {\bf 168733} & 6.3526 & 6.3000 \\
{\bf 21699} & 2.4920 & 2.4761 & {\bf 56455} & 2.0583 & 2.0580 & {\bf 125823} & 8.8166 & 8.8171 & {\bf 170000} & 1.7166 & 1.7165 \\
{\bf 22470} & 1.9289 & 1.9300 & {\bf 59256} & 0.9812 & 0.9430 & {\bf 128898} & 8.9051 & 0.0047 & {\bf 175362} & 3.6741 & 3.6740 \\
{\bf 22920} & 3.9471 & 3.9474 & {\bf 66624} & 1.9983 & 1.9984 & {\bf 130158} & 4.4142 & 4.4136 & {\bf 182255} & 1.2622 & 1.2600 \\
{\bf 23408} & 20.3234 & 10.2880 & {\bf 72968} & 5.6520 & 11.3050 & {\bf 131120} & 1.5687 & 1.5690 & {\bf 183056} & 0.7450 & 0.6870 \\
{\bf 25267} & 1.2099 & 1.2094 & {\bf 73340} & 2.6676 & 2.6675 & {\bf 133880} & 0.8775 & 0.8775 & {\bf 183806} & 3.1197 & 2.9210 \\
{\bf 25823} & 7.2267 & 7.2274 & {\bf 74067} & 3.1129 & 3.1130 & {\bf 134759} & 1.5566 & 3.0993 & {\bf 196178} & 1.0041 & 1.0060 \\
{\bf 26961} & 1.5274 & 1.5274 & {\bf 74521} & 7.0519 & 6.9071 & {\bf 137909} & 18.2751 & 18.4870 & {\bf 196502} & 10.1376 & 20.2750 \\
{\bf 27309} & 1.5689 & 1.5690 & {\bf 74560} & 1.5509 & 1.5511 & {\bf 138764} & 1.2586 & 1.2588 & {\bf 201433} & 1.1260 & 1.4289 \\
{\bf 28843} & 1.3738 & 1.3740 & {\bf 77653} & 1.4878 & 1.4878 & {\bf 138769} & 2.0895 & 2.0894 & {\bf 201834} & 3.5351 & 3.5354 \\
{\bf 29009} & 3.8002 & 3.8200 & {\bf 79158} & 1.9174 & 3.8340 & {\bf 140160} & 1.5958 & 1.5958 & {\bf 209515} & 2.3870 & 2.3880 \\
{\bf 29305} & 2.9432 & 2.9500 & {\bf 82984} & 6.0878 & & {\bf 140728} & 1.2956 & 1.3049 & {\bf 220825} & 1.4149 & 1.4150 \\
{\bf 32549} & 4.6396 & 4.6397 & {\bf 92664} & 1.6731 & 1.6680 & {\bf 142096} & 6.6411 & 3.3136 & {\bf 221006} & 2.3148 & 2.3148 \\
{\bf 32650} & 2.7350 & 2.7348 & {\bf 93030} & 2.2031 & 2.2027 & {\bf 142990} & 0.9789 & 0.9789 & {\bf 223640} & 3.7350 & 3.7352 \\
\hline
\end{tabular}
\end{center}
\end{table*}

\subsection{Variable stars} \label{variable_stars}

In total, we identified 84 variable stars, from which 83 have periods in the International Variable Star Index \citep[VSX,][]{2006SASS...25...47W} of the American Association of Variable Star Observers and the compilations of \citet{2017MNRAS.468.2745N} and \citet{2019A&A...622A.199J}. Example phased light curves of three sample stars (HD 11502, HD 25823, and HD	74067) are shown in Figure \ref{phased_light_curves}. The different light curves characteristics relating to the spot distribution on the surfaces are nicely visible.

In Fig. \ref{comparison_known_variables}, we present a comparison of the here derived periods to the literature periods of the stars listed in Table \ref{found_periods}. The typical period error amounts to 0.0001\,d. Because in orientations where two spots come into view during a star's rotation cycle, a significant number of ACV variables exhibit double-waved light curves \citep{2016AJ....152..104H}. Therefore, a twice longer (or shorter) rotation period is sometimes possible, in particular for objects with very small amplitudes and/or significant scatter in their light curves. We emphasise that the given period values correspond to the periods with the highest FAP values; nevertheless, the most prominent periods may sometimes represent harmonics of the true rotation period, as becomes obvious from Fig. \ref{comparison_known_variables}. In ambiguous cases, the true period has to be determined by other means such as radial velocity studies.

One object (HD 82984) is listed as ''CST:'' in the VSX, that is, as a non-variable (constant) star, which was formerly suspected to be a variable. This classification goes back to \citet{1977AcA....27..365J}, who did not discuss HD 82984 in more detail. No further references to a photometric time-series study of this object were found in the literature; we are therefore confident that the here presented solution is the first published period of this star.

There are three obvious outliers in Fig. \ref{comparison_known_variables}, which are discussed in more detail below.

{\it HD 183056:} \citet{1974A&A....35..381W} reported variability of the H$\beta$-line profile on the basis of 49 scans taken during a time span of 188 min. Later on, \citet{1974PhDT........61W} reported a possible period of 0.68674\,d for this star, whereas \citet{1993A&AS..101..393A} concluded that if variability is present, the amplitude has to be below 0.02\,mag. HD 183056 was included in the {\it Hipparcos}-based study of \citet{1998A&AS..132...93A}; however, no period solution was presented.

{\it HD 183806:} \citet{1985A&AS...60...17M} measured this star in Str{\"o}mgren $uvby$ filters and reported a possible period of 2.9213\,d. No other period information on this star seems to have been published. The here derived SMEI period of 3.1197\,d is slightly longer than the published one.

{\it HD 201433:} This is a single-lined spectroscopic triple system consisting of a massive Slowly Pulsating B (SPB) star orbited by two low-mass stars with periods of about 3.31 and 154\,d, respectively \citep{2017A&A...603A..13K}. The amplitude spectrum of the SMEI data is clearly indicative of multiperodicity due to pulsations. The here detected periods (not all listed in Table
\ref{found_periods}) are compatible with those listed in Table 4 of \citet{2017A&A...603A..13K}.

\begin{figure}
\begin{center}
\includegraphics[width=0.49\textwidth]{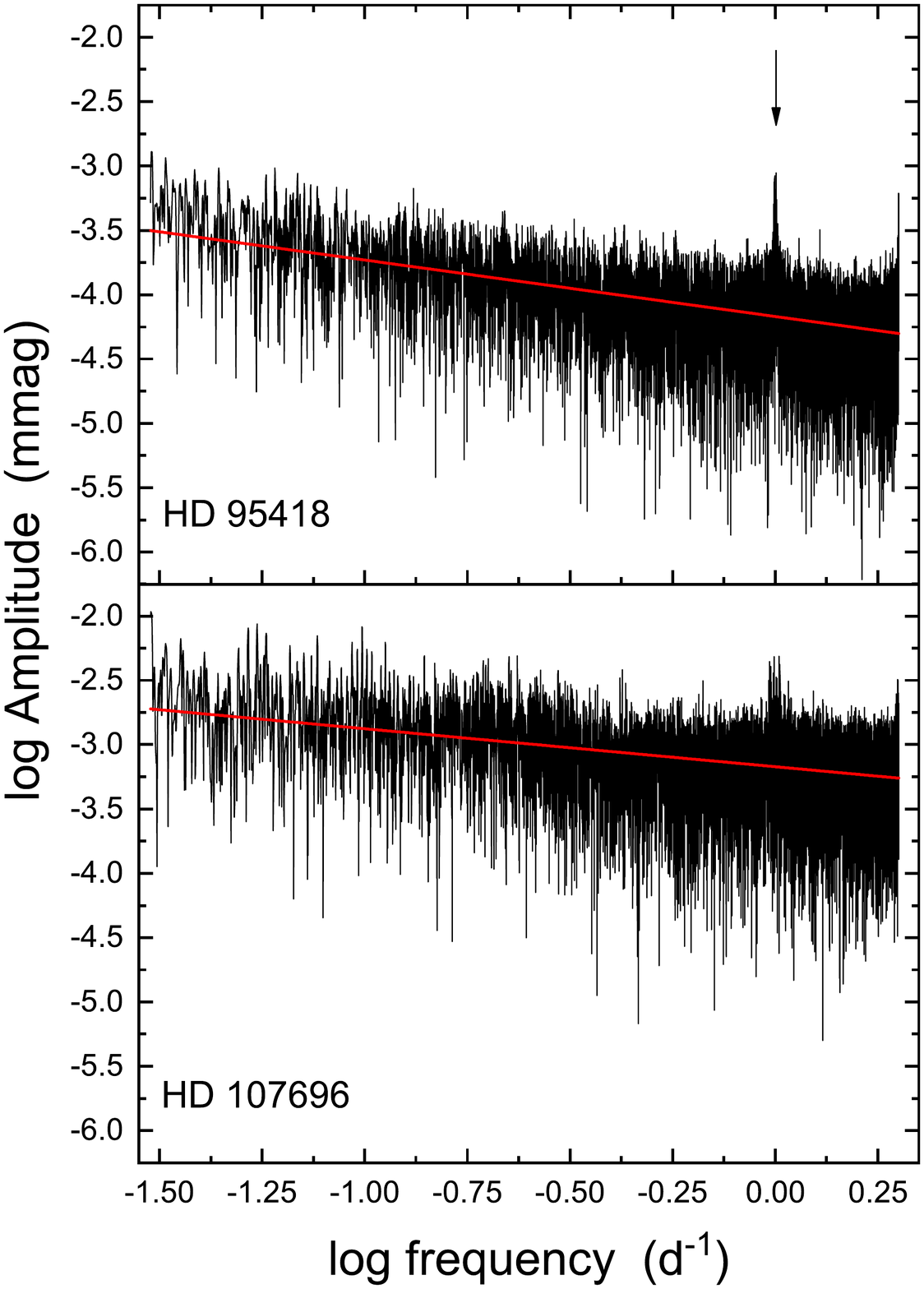}
\caption{Logarithmic amplitude spectra of the SMEI light curves of HD 95418 (upper panel) and HD 107696 (lower panel). Also shown are the fits with the parameters listed in Table \ref{result_constant}. In both amplitude spectra, the remains of the one-day frequency is clearly visible.}
	\label{spectrum_noise}
\end{center}
\end{figure}

All other stars lie well on the correlations presented in Fig. \ref{comparison_known_variables}, which lends confidence in our time-series analysis.

\begin{table*}
\begin{center}
\caption{Results from fitting the amplitude spectra (Fig. \ref{spectrum_noise}) of the apparently constant stars in the form $\log A = a + b \log f$, where $A$ is the amplitude and $f$ the frequency. Throughout the sample, the errors relating to the parameters $a$ and $b$ amount to 0.002 and 0.005, respectively}
\label{result_constant}
\begin{tabular}{ccc|ccc|ccc|ccc}
  \hline
{\bf HD} & $a$ & $b$ & {\bf HD} & $a$ & $b$ & {\bf HD} & $a$ & $b$ & {\bf HD} & $a$ & $b$ \\
\hline
{\bf 1280} & $-$3.944 & $-$0.461 & {\bf 64740} & $-$3.976 & $-$0.414 & {\bf 108945} & $-$3.723 & $-$0.448 & {\bf 181615} & $-$3.722 & $-$0.550 \\
{\bf 2054} & $-$3.765 & $-$0.445 & {\bf 68351} & $-$3.563 & $-$0.458 & {\bf 115735} & $-$3.901 & $-$0.443 & {\bf 182568} & $-$3.824 & $-$0.454 \\
{\bf 4853} & $-$3.840 & $-$0.492 & {\bf 68601} & $-$3.990 & $-$0.475 & {\bf 116458} & $-$3.794 & $-$0.335 & {\bf 185872} & $-$3.468 & $-$0.439 \\
{\bf 5737} & $-$3.837 & $-$0.431 & {\bf 71046} & $-$3.893 & $-$0.317 & {\bf 116656} & $-$4.100 & $-$0.466 & {\bf 187474} & $-$3.634 & $-$0.431 \\
{\bf 6397} & $-$3.678 & $-$0.471 & {\bf 77350} & $-$3.612 & $-$0.446 & {\bf 120640} & $-$3.454 & $-$0.419 & {\bf 188041} & $-$3.587 & $-$0.380 \\
{\bf 11753} & $-$3.806 & $-$0.424 & {\bf 78556} & $-$3.526 & $-$0.404 & {\bf 120709} & $-$3.841 & $-$0.490 & {\bf 189178} & $-$3.822 & $-$0.461 \\
{\bf 18519} & $-$3.950 & $-$0.479 & {\bf 78702} & $-$3.677 & $-$0.370 & {\bf 123255} & $-$3.709 & $-$0.478 & {\bf 196524} & $-$3.990 & $-$0.449 \\
{\bf 19400} & $-$3.798 & $-$0.331 & {\bf 79416} & $-$3.638 & $-$0.452 & {\bf 124425} & $-$3.755 & $-$0.436 & {\bf 198667} & $-$3.708 & $-$0.407 \\
{\bf 23850} & $-$3.769 & $-$0.449 & {\bf 81188} & $-$4.092 & $-$0.361 & {\bf 128974} & $-$3.472 & $-$0.464 & {\bf 201601} & $-$3.897 & $-$0.442 \\
{\bf 28319} & $-$3.858 & $-$0.451 & {\bf 90264} & $-$3.829 & $-$0.370 & {\bf 130109} & $-$3.914 & $-$0.445 & {\bf 202444} & $-$4.050 & $-$0.437 \\
{\bf 30612} & $-$3.717 & $-$0.333 & {\bf 90972} & $-$3.534 & $-$0.428 & {\bf 135382} & $-$3.918 & $-$0.314 & {\bf 202627} & $-$3.854 & $-$0.461 \\
{\bf 34968} & $-$3.920 & $-$0.479 & {\bf 92728} & $-$3.666 & $-$0.374 & {\bf 136933} & $-$3.498 & $-$0.449 & {\bf 202671} & $-$3.608 & $-$0.430 \\
{\bf 35039} & $-$3.925 & $-$0.437 & {\bf 94334} & $-$3.934 & $-$0.417 & {\bf 143699} & $-$3.668 & $-$0.440 & {\bf 203006} & $-$3.758 & $-$0.448 \\
{\bf 35497} & $-$4.037 & $-$0.463 & {\bf 95418} & $-$4.169 & $-$0.439 & {\bf 148330} & $-$3.917 & $-$0.456 & {\bf 203585} & $-$3.453 & $-$0.450 \\
{\bf 38104} & $-$3.872 & $-$0.435 & {\bf 96097} & $-$3.675 & $-$0.552 & {\bf 152127} & $-$3.690 & $-$0.414 & {\bf 205637} & $-$3.630 & $-$0.424 \\
{\bf 42509} & $-$3.413 & $-$0.468 & {\bf 97633} & $-$3.961 & $-$0.487 & {\bf 157740} & $-$3.480 & $-$0.424 & {\bf 206742} & $-$3.916 & $-$0.504 \\
{\bf 55719} & $-$3.932 & $-$0.444 & {\bf 98664} & $-$3.674 & $-$0.557 & {\bf 169467} & $-$3.813 & $-$0.439 & {\bf 215766} & $-$3.213 & $-$0.407 \\
{\bf 59635} & $-$3.743 & $-$0.472 & {\bf 101189} & $-$3.607 & $-$0.378 & {\bf 172728} & $-$3.783 & $-$0.476 & {\bf 221760} & $-$3.768 & $-$0.445 \\
{\bf 61110} & $-$3.828 & $-$0.472 & {\bf 106661} & $-$3.687 & $-$0.510 & {\bf 175156} & $-$3.736 & $-$0.403 &  & & \\
{\bf 61641} & $-$3.706 & $-$0.429 & {\bf 107696} & $-$3.172 & $-$0.295 & {\bf 177003} & $-$3.779 & $-$0.400 &  & & \\
{\bf 64486} & $-$3.980 & $-$0.505 & {\bf 108283} & $-$3.723 & $-$0.473 & {\bf 177756} & $-$4.003 & $-$0.470 &  & & \\
\hline
\end{tabular}
\end{center}
\end{table*}

\subsection{Apparently constant stars} \label{constant_stars}

Here, we define ''constant'' as light curves whose amplitude spectra show no significant peaks as defined in Section \ref{tsa}. As first step, we describe the characteristics of the amplitude spectra in more detail.

The noise in the amplitude spectra can be well described as 1/f or ''flicker'' noise. Sometimes it also referred to as ''pink'' noise \citep{1997atsa.book.....S}. It can be described as a linear law by plotting the logarithm of the frequency ($f$) versus the logarithm of the amplitude ($A$).

\begin{equation}
\log A = a + b \log f.
\end{equation}

The $a$ and $b$ values of the 81 amplitude spectra without any significant peaks are listed in Table \ref{result_constant}. The errors of the parameters $a$ and $b$ amount to 0.002 and 0.005, respectively, throughout the sample, which implies that the estimates are very robust. The mean slope for the complete sample is $-$0.44(5), which means that the noise characteristics are not dependent on the individual data sets. The level of the noise characterised by the intercept depends on the number of data points and the magnitude of the stars (photon statistics).

In Fig. \ref{spectrum_noise}, we present example amplitude spectra of two stars showing, respectively, a low (HD 95418, upper panel) and a high (HD 107696, lower panel) noise level in the investigated frequency domain. The fits are shown as straight lines. In both amplitude spectra, the remains of the one-day frequency are clearly visible.

\begin{figure}
\begin{center}
\includegraphics[width=0.49\textwidth]{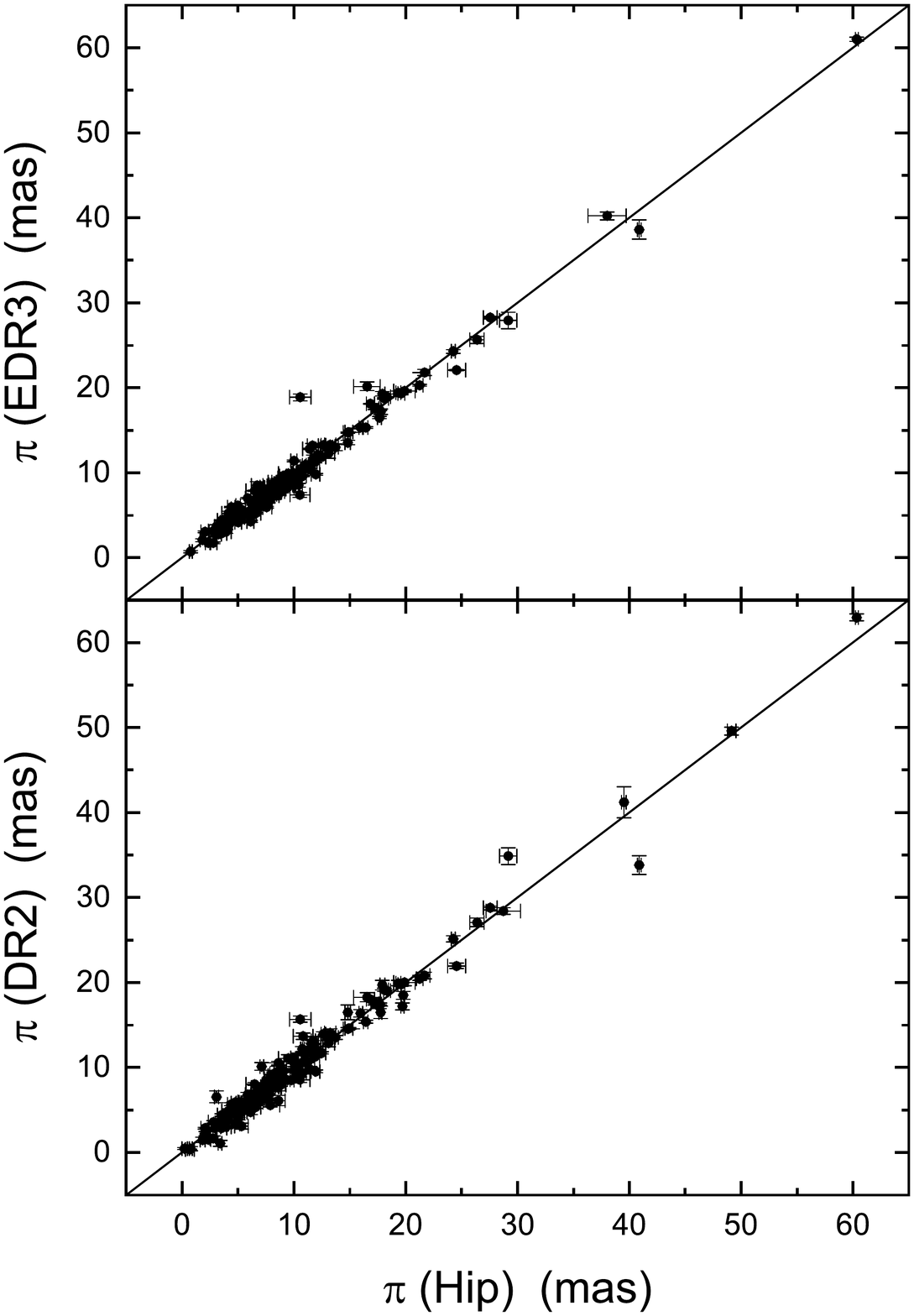}
\caption{Comparison of the parallaxes from the {\it Hipparcos} catalogue to the {\it Gaia} EDR3 parallaxes (upper panel) and the {\it Gaia} DR2 parallaxes (lower panel). Several obvious outliers are visible in both plots.}
	\label{comparison_Hip_Gaia}
\end{center}
\end{figure}

From the 81 stars without any significant signals in SMEI data, 18 stars have periods in the literature and another 10 stars are flagged as variables or variable star candidates in the VSX. In the following, we discuss these objects in more detail.

In the case of four stars, the literature period exceeds 30 days, which may explain their non-detection in SMEI data: HD 116458 (149.25373\,d), HD 181615 (137.9343\,d), HD 187474 (2300\,d), and HD 188041 (224.5\,d).

\citet{2012MNRAS.420..757W} identified low amplitude variations on the basis of STEREO data in HD 6397, HD 68351, and HD 77350.

One of the most valuable sources for the study of bright ACV variables are the data of the {\it Hipparcos} mission. \citet{1998A&AS..132...93A} and \citet{1998A&AS..133....1P} reported low amplitude variability in HD 73340, HD 79416, HD 201601, HD 202671, HD 203006, HD 205637, and HD 221760.

{\it HD 5737:} the amplitude of this He-weak object amounts to only a few mmags in the Str{\"o}mgren photometric system \citep{1994A&A...281...73M}.

{\it HD 23850:} this star was found constant in light by \citet{2012MNRAS.420..757W}.

{\it HD 28319:} this is the binary system $\theta^2$ Tauri, the brightest member of the Hyades, which consists of at least two components showing $\delta$ Scuti-type pulsations \citep{2006AJ....131.2643A}. The existence of rotational variability was never confirmed.

{\it HD 35039:} \citet{1985MNRAS.214..559B} found first evidence for line-profile variations in this star using a limited set of observations. Later on, \citet{2001A&A...380..277L} classified HD 35039 as a SPB star showing multiperiodic radial velocity variations. However, it is a He-rich CP star according to \citet{2013AstBu..68..300R}. This object deserves a new thorough analysis of all spectroscopic data to shed more light on its nature.

{\it HD 68601:} this star was found variable by \citet{1977AcA....27..365J}, who did not discuss its characteristics in more detail. No other variability study of this object was found in the literature.

{\it HD 107696:} this star was part of the long-term photometric survey of variable stars at ESO \citep{1991A&AS...87..481M}. However, no detailed results were published on this object by the consortium.

{\it HD 108945:} this star was analysed in more detail with photometric and spectroscopic data by \citet{2019MNRAS.485.4247P}, who confirmed the existence of rotational light variability but were unable to find any significant frequencies with a semi-amplitude exceeding 0.2 mmag in the high-frequency range.

{\it HD 115735:} this is a well-studied ACV variable \citep{1984BAICz..35..294Z,1994A&A...283..932Z}.

{\it HD 124425:} no time-series analysis was found in the literature.

{\it HD 148330:} \citet{1983IBVS.2366....1Z} identified radial velocity
variations and photometric variability with an amplitude of 17 mmag in the $g_2$ filter \citep{2005A&A...441..631P}.

{\it HD 215766:} this star was found constant by \citet{2012MNRAS.420..757W}.

Upper limits of variability were derived for 50 out of 165 objects ($\sim$30\%) for which no entry about variability exists in the VSX and the literature. The question arises which parameters define the boundary conditions for the occurrence of ACV variability, that is, the presence of spots of sufficient size or contrast on the surface that lead to observable photometric variations.

\citet{2013MNRAS.431.1883J} analysed the relationship between size and surface coverage of starspots on magnetically-active low-mass stars. Employing different filling factors, characteristic scalelengths, and spot distributions in their simulations, they were able to show that small light-curve amplitudes in magnetically active M dwarfs can be explained by a starspot model consisting of large filling factors of dark spots with a random distribution on the stellar surface. To adjust this model to higher mass objects and use it for a comparison with the observations, a much larger sample size of non-variable CP2/4 stars is needed. Although a huge amount of high-quality photometric light curves are available today, such a project -- although being of great interest -- is not easy to implement \citep{2020svos.conf...19P}.

\section{Astrophysical parameters and the Hertzsprung-Russell-diagram} \label{ap_hrd}

In this section, we describe the sources and methods used to create the Hertzsprung-Russell diagram for our target star sample.

{\it Photometry:} photometric data of the Johnson $UBV$ and Geneva 7-colour systems were taken from the General Catalogue of Photometric data\footnote{\url{http://gcpd.physics.muni.cz/}} and the All-Sky Compiled Catalogue of 2.5 million stars \citep[ASCC,][]{2001KFNT...17..409K}, whereas the Str{\"o}mgren-Crawford $uvby\beta$ indices were gleaned from the catalogue by \citet{2015A&A...580A..23P}.

No Johnson $(U-B)$ colours are available for HD 3980, HD 18519, HD 54118, HD 59635, HD 166596, HD 168733, and HD 183806. The full set of Str{\"o}mgren-Crawford $uvby\beta$ indices was measured for all objects, except for HD 4853. Geneva 7-colour photometry is not available for HD 21699, HD 23408, HD 108662, HD 108945, HD 172728, HD 177003, HD 182568, HD 185872, and HD 189178.
 
{\it Reddening:} as was shown by \citet{2008A&A...491..545N}, the commonly employed dereddening procedures published by \citet{1993A&A...268..653N} are also applicable to CP2/4 stars. They were therefore used for the reddening estimation of all objects except HD 4853, in which case we used the reddening map\footnote{\url{http://argonaut.skymaps.info/}} by \citet{2019ApJ...887...93G}, which is based on Gaia parallaxes and stellar photometry from Pan-STARRS 1 and 2MASS. Because this star is located in the solar vicinity (distance of 82\,pc), the reddening is negligible.

For the calibrations of the different photometric systems, we used the following relations \citep{2006A&A...458..293P}:
\begin{equation}
A_V = 3.1E(B-V) = 4.3E(b-y) = 4.95E(B2-V1).
\end{equation}
A value of 0.01\,mag in $E(b-y)$ was adopted for all objects.

{\it Bolometric correction:} we used the correlation given by \citet{2008A&A...491..545N}, which was tailored for the use with CP stars.

{\it Luminosity:} today, regarding bright stars, we are in the lucky position to have parallaxes available from both the {\it Hipparcos} \citep{2007A&A...474..653V} and the {\it Gaia} \citep{2018A&A...616A...1G,2018A&A...616A...2L,2020arXiv201201533G} space missions. Because our sample consists of stars between magnitudes 1.5 and
7 (Fig. \ref{histograms}), we checked the consistency of both data sources. Only HD 36485 has no parallax listed in the {\it Hipparcos} catalogue, whereas measurements for 11 stars (HD 29305, HD 35497, HD 77653, HD 81188, HD 93030, HD 96097, HD 103192, HD 116656, HD 170000, HD 196524, and HD 203585) and 16 stars (HD 29305, HD 35497, HD 40312, HD 77653, HD 81188, HD 93030, HD 96097, HD 97633, HD 103192, HD 112185, HD 116656, HD 134759, HD 170000, HD 196524, HD 202444, and HD 203585) are missing in the {\it Gaia} DR2 and EDR3, respectively.

In Fig. \ref{comparison_Hip_Gaia}, we present the comparison of the parallaxes from the {\it Hipparcos} and {\it Gaia} data releases. In general, the agreement is very good, but some outliers are present. We have checked the available literature to search for reasons to account for these discrepancies. Most likely, binarity or circumstellar envelopes are at the root of the observed outlying positions, but it is out of the scope of this paper to investigate this issue in detail and to decide which of the parallax sources is the most reliable one.

Parallaxes were used in the sequence EDR3, DR2, and Hipparcos, depending on their availability. The parallax values were used to calculate the final luminosities listed in Table \ref{table_master2}; a full error propagation was applied.

\begin{figure}
\begin{center}
\includegraphics[trim = 0mm 0mm 10mm 125mm, clip, width=0.49\textwidth]{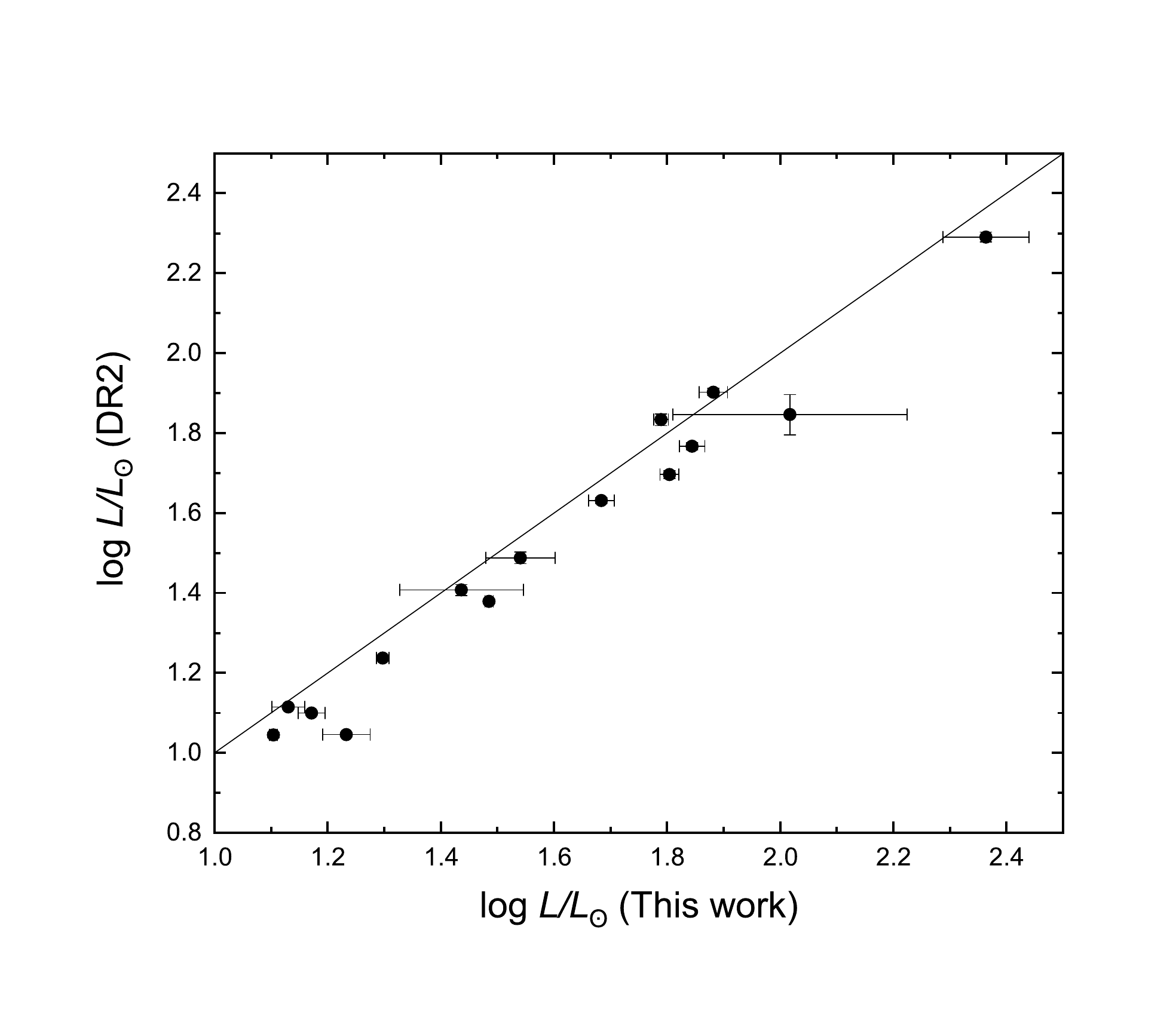}
\caption{Comparison of the luminosities derived in this work and the {\it Gaia} DR2 catalogue. The latter appear systematically too low.}
	\label{comparison_luminosities}
\end{center}
\end{figure}

$Gaia$ DR2 also includes luminosities for objects with effective temperatures less than 10\,000\,K \citep{2020A&A...634A..18C}. Corresponding luminosities were obtained for 15 stars from our list. A comparison of the {\it Gaia} DR2 luminosities to our values is provided in Figure \ref{comparison_luminosities}, from which it is obvious that the {\it Gaia} DR2 luminosities are systematically too low and should be used with caution for CP2/4 stars.

\begin{figure}
\begin{center}
\includegraphics[width=0.49\textwidth]{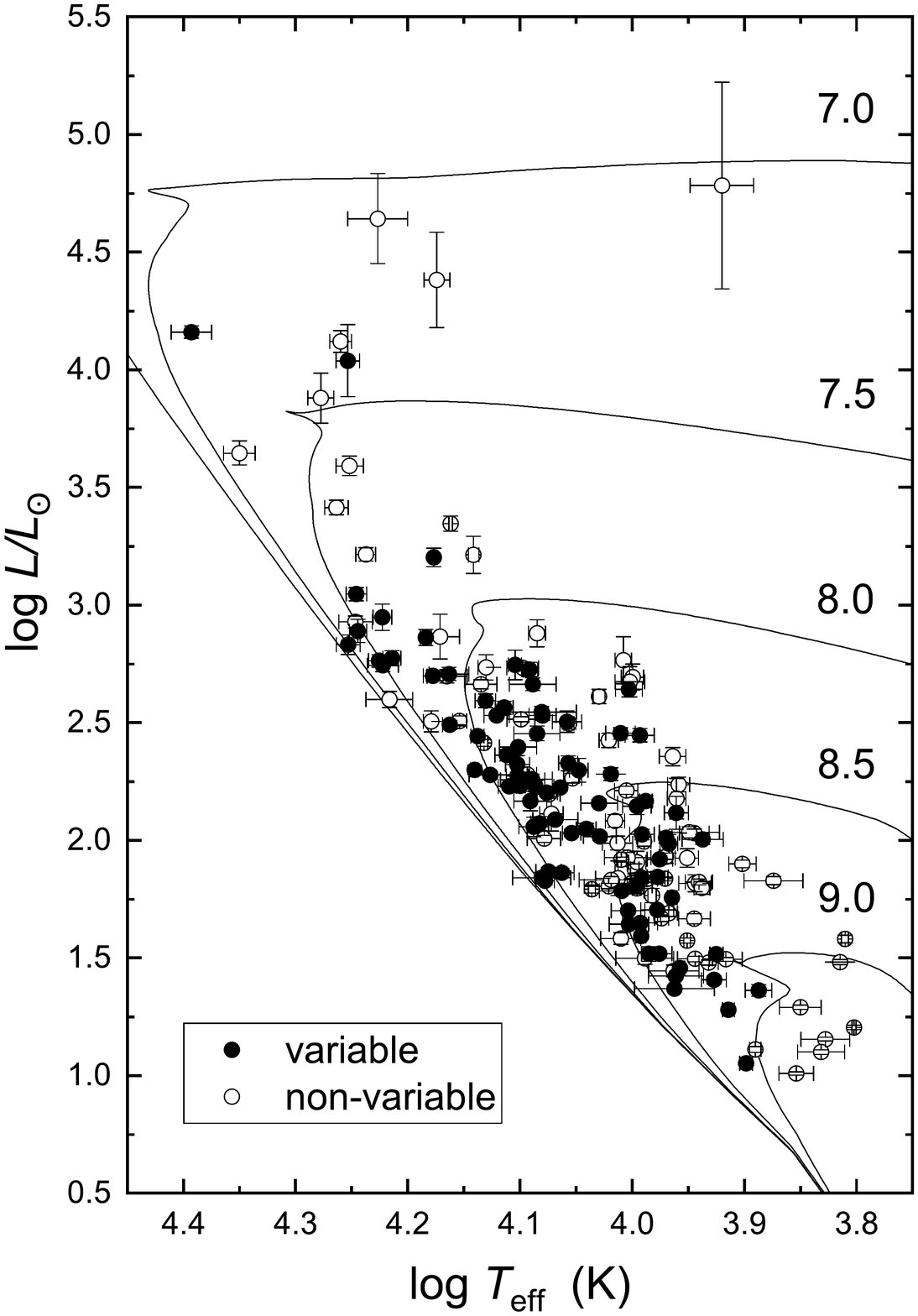}
\caption{Positions of our target stars in the Hertzsprung-Russell diagram, which is provided in the form of $\log T_\mathrm{eff}$ versus $\log L/L_\odot$ (Table \ref{table_master1}). Also shown are stellar evolutionary models for the indicated logarithmic ages by \citet{2012A&A...537A.146E}.}
	\label{hrd}
\end{center}
\end{figure}

{\it Effective temperature:} if available, data from the Johnson $UBV$, Geneva 7-colour, and Str{\"o}mgren-Crawford $uvby\beta$ photometric systems were used. \citet{2008A&A...491..545N} introduced calibrations for CP stars using individual corrections for the temperature domain and the CP subclass, which are summarised in their Table 2. We here follow their approach. For the derivation of the final effective temperatures, all calibrated values were averaged and the standard deviations were calculated.

{\it Mass} and {\it Evolutionary Phase:} to estimate these parameters, we employed the Stellar Isochrone Fitting Tool\footnote{\url{https://github.com/Johaney-s/StIFT}}, which builds on the methodology of \citet{2010MNRAS.401..695M} in estimating the mass, age, radius, and evolutionary phase of a star according to its effective temperature and luminosity. The tool automatically searches for similar data in models based on evolutionary tracks and selects the four grid points closest to the input value in the Hertzsprung-Russell diagram. From these grid points, the output parameters are obtained by repetitive linear interpolation.

The final astrophysical parameters are listed in Table \ref{table_master1}. In Fig. \ref{histograms} (lower panel), the distribution of the effective temperatures is shown. Most stars have values between 9\,000 and 13\,000\,K, which corresponds to spectral types between A2 and B8, in agreement with the peak distribution of CP stars. As expected, there is also an extension to cooler and hotter temperatures that covers the whole range of the CP star phenomenon.

In Fig. \ref{hrd}, the Hertzsprung-Russell diagram is presented together with non-rotating isochrones from \citet{2012A&A...537A.146E} for [Z]\,=\,0.014. 
The choice of the metallicity rests on the analysis presented in Sect. 4.2.2. of \citet{2020A&A...640A..40H}. It is based on published individual abundances for stars of
our target star sample. No object is located below the zero-age main sequence, but several stars are close to the terminal-age main sequence, which implies that they are past the core-hydrogen burning stage and do not belong to luminosity class V. The most significantly deviating object is HD 68601. It was classified as ApSr on the basis of S2/68 UV spectra by \citet{1978A&AS...33...15C}. Later on, it was classified as A5\,Ib by \citet{1989ApJS...70..623G}, which is compatible with its location in the Hertzsprung-Russell diagram as shown in Fig. \ref{hrd}. The $\Delta$a value, which provides a measure of peculiarity by an estimation of the 5200\,\AA\ flux depression, is slightly positive \citep[+10\,mmag,][]{1998A&AS..130..455V}. Unfortunately, no detailed elemental abundance studies are available in the literature. Spectroscopic studies of this interesting and apparently evolved star are encouraged.

\begin{figure}
\begin{center}
\includegraphics[width=0.49\textwidth]{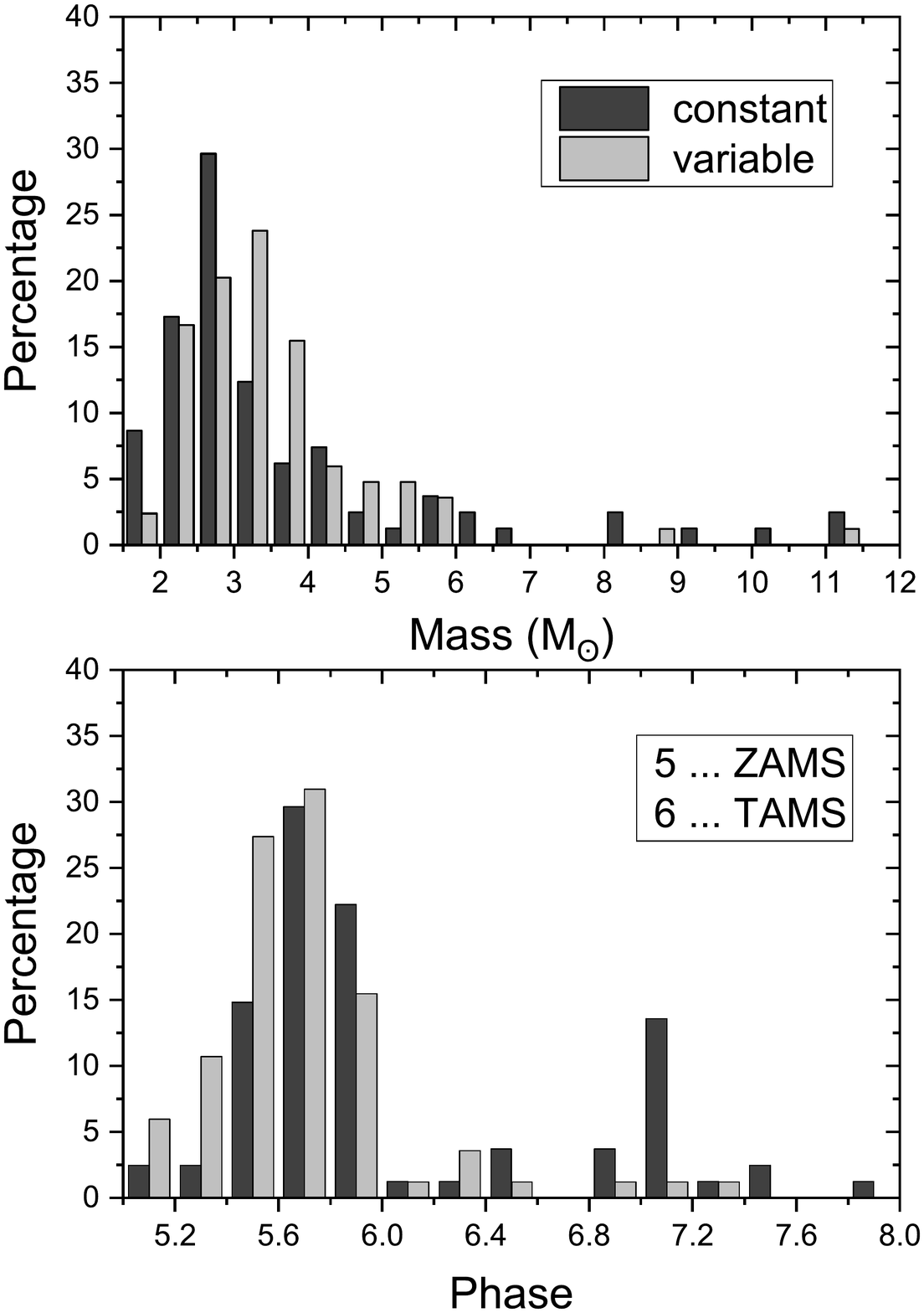}
\caption{Histograms of the masses and evolutionary phases of the constant and variable CP2/4 stars. The coding of the evolutionary phase follows the isochrone grid of \citet{2012MNRAS.427..127B}.}
	\label{mass_phase}
\end{center}
\end{figure}

Figure \ref{mass_phase} shows the mass and evolutionary status of the sample divided into variable (Table\,\ref{found_periods}) and constant (Table\,\ref{result_constant}) stars. The coding for the evolutionary status was taken from the isochrone grid by \citet{2012MNRAS.427..127B}. The mass distribution peaks at about three solar masses, as expected from the effective temperature distribution (Figure \ref{histograms}). Interestingly, variable stars show significantly higher masses than constant ones. This is expected because high-mass CP2/4 stars usually have He and Si spots, which are known to produce higher-amplitude variations than Sr, Cr, or Eu spots.

The age distribution is similar to the distributions of the fainter ACV variables investigated by \citet{2015A&A...581A.138B,2016AJ....152..104H,2020MNRAS.493.3293B}. There is a conspicuous lack of young stars close to or at the zero-age main sequence. Almost all objects have ages between 100 Myr and 1 Gyr (Figure \ref{hrd}), which still places them on the main sequence but already significantly above the zero-age main sequence line. Variable stars seem to be in a more advanced evolutionary status than the constant ones. This information is important for theories dealing with the origin of the observed magnetic fields in these objects. Over the last years, a body of evidence has been built up which strongly favours the fossil field theory, which postulates that the magnetic
fields are relics of the ''frozen-in'' interstellar magnetic field \citep{2013MNRAS.428.2789B}. Our results imply that spots on the stellar surface, which are thought to be coupled to the magnetic field strength, only appear after the star has spent some time on the main-sequence. From studies of open clusters \citep{2003A&A...403..645B,2003A&A...402..247P,2008A&A...481..465L}, it has been established that young CP2/4 stars exist, although they seem to be rare. Young CP2/4 stars are present in the large sample of stars published by \citet{2020A&A...640A..40H}, although they are conspicuously underrepresented. The reasons for the observed uneven distribution of CP2/4 stars are still unresolved and a matter of debate.

\begin{figure}
\begin{center}
\includegraphics[width=0.49\textwidth,bb=0 0 432 519]{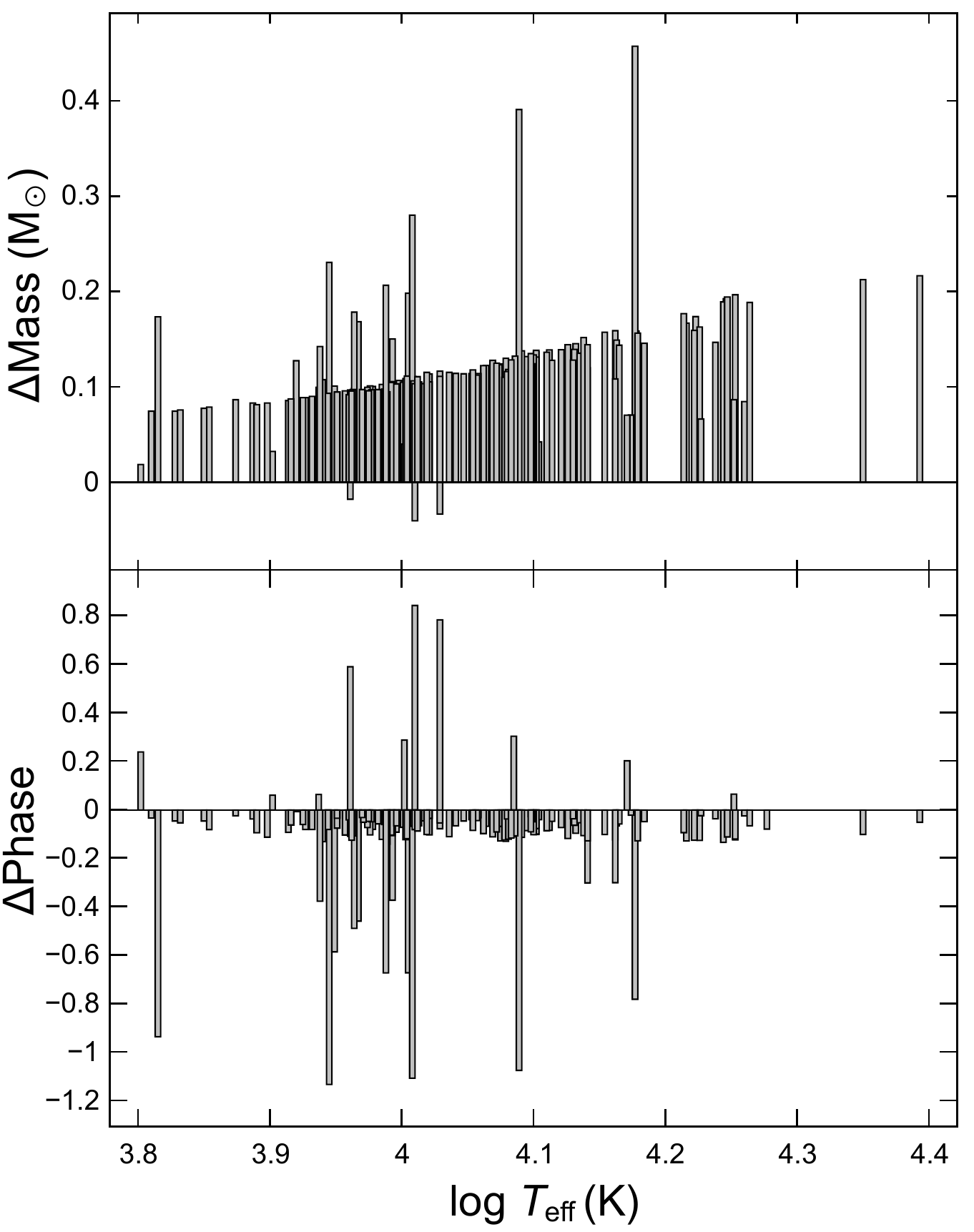}
\caption{Differences of mass and evolutionary phase estimated from tracks with [Z]\,=\,0.020 and [Z]\,=\,0.014.}
	\label{parameters_by_z}
\end{center}
\end{figure}

Finally, we checked the influence of the metallicity on the mass and evolutionary phase distribution. For this, we used an identical isochrone grid for [Z]\,=\,0.020 and calibrated the target star sample as described above. In Fig. \ref{parameters_by_z}, the differences of mass and evolutionary phase in the sense ''0.020 $-$ 0.014'' is shown. Derived masses are between 0.1 and 0.2 $M_\odot$ larger for the higher metallicity, which is well below 5\% for the investigated mass range (Fig. \ref{mass_phase}). The evolutionary phase is constantly about 0.1 smaller in the sense that higher metallicity results in younger ages. Again, these absolute differences are negligible when compared to the observational uncertainties and possible effects such as unknown binarity. The bin size used for the analysis presented in Fig. \ref{mass_phase} is 0.2 for the evolutionary phase. About 10\% of the stars from our sample do not follow the general trend. These are the objects situated close to, or slightly above, the terminal-age main sequence. The observed sudden ''jump'' with the isochrone's metallicity is due to the corresponding shift in luminosity for the same effective temperature, which decides whether these objects are still calibrated as core hydrogen-burning objects or not.

\section{Conclusions} \label{conclusions}

Using SMEI data, we investigated the light curves of 165 CP2 and CP4 stars and candidates selected from the catalogue by \citet{RM09}. Because of the presence of strong instrumental trends, a basic cleaning algorithm was introduced, and instrumentally-introduced frequencies were identified and subtracted from the data sets in an automated way.

The time series analysis was performed using the Generalized Lomb-Scargle algorithm and the Phase-Dispersion Method. Because both methods complement each other and are sensitive to the different light curve characteristics and instrumental effects, this approach was found to work well with SMEI data and to derive the most reliable results.

Apart from performing thorough time-series analyses, it is also important to establish criteria defining constancy. We found that the noise of the amplitude spectra can be well described as flicker (or pink) noise, which, when plotting the logarithm of the frequency versus the logarithm of the amplitude, is characterised by a linear law that describes the mean noise level over the investigated frequency domain. Corresponding parameters for all apparently constant stars were presented.

In total, we processed and analysed 165 individual light curves from which 84 show variability in the accuracy limit of the SMEI data. We compared the derived periods with published literature values and found an excellent agreement, which proves that SMEI data are well suited to the long-term study of ACV variables.

As last step, we used calibrations specifically developed for CP stars as well as {\it Hipparcos} and {\it Gaia} data to estimate the effective temperatures and luminosities for our target star sample. Masses and evolutionary stages were calibrated using an appropriate isochrone grid. The derived astrophysical parameters provide a coherent picture of ACV variables being concentrated towards the end of their main-sequence lifetime. Our mass estimates support that He and Si peculiarities -- preferentially found in the hotter, and thus more massive, CP stars -- produce larger spots or spots of higher contrast.

SMEI data fill an important gap in time in the observation of bright ACV variables, for which follow-up spectroscopic and spectropolarimetric observations are comparatively easy to achieve. The here presented results are therefore important for the future analysis and, in particular, the study of the long-term behaviour of CP2/4 star light curves, and will benefit future studies of these fascinating objects.

\section*{Acknowledgements}

This paper is dedicated to Leo Warzecha who died during its preparation.
EP acknowledges support by the Erasmus+ programme of the European Union under grant number 2020-1-CZ01-KA203-078200. This work has also made use of data from the European Space Agency (ESA) mission {\it Gaia}, processed by the {\it Gaia} Data Processing and Analysis Consortium (DPAC) \url{https://www.cosmos.esa.int/web/gaia/dpac/consortium}). Funding for the DPAC has been provided by national institutions, in particular the institutions participating in the {\it Gaia} Multilateral Agreement. This research has made use of the SIMBAD database, operated at CDS, Strasbourg, France.

\section*{Data availability}

The data underlying this article will be shared on reasonable request to the corresponding author.

\bibliographystyle{mnras}
\bibliography{SMEI}

\appendix

\section{Essential data for our sample stars}

Table \ref{table_master1} lists essential data for our sample stars. It is organised as follows:

\begin{itemize}
\item Column 1: HD number.
\item Column 2: Identifier from \citet{RM09}.
\item Column 3: Right ascension (J2000). Positional information was taken from GAIA DR2 \citep{2018A&A...616A...1G,2018A&A...616A...2L}.
\item Column 4: Declination (J2000).
\item Column 5: Spectral classification from \citet{RM09}.
\item Column 6: Mean $V$ magnitude.
\item Column 7: Mean $V$ magnitude error.
\item Column 8: Parallax (Hipparcos).
\item Column 9: Parallax error.
\item Column 11: Parallax (Gaia DR2).
\item Column 11: Parallax error.
\item Column 12: Parallax (Gaia EDR3).
\item Column 13: Parallax error.
\item Column 14: Absorption in the $V$ band.
\item Column 15: Mean effective temperature.
\item Column 16: Mean effective temperature error.
\item Column 17: Luminosity.
\item Column 18: Luminosity error.
\item Column 19: Mass.
\item Column 20: Evolutionary Phase.
\end{itemize}

\setcounter{table}{0}  
\begin{table*}
\caption{Essential data for our sample stars, sorted by increasing right ascension. The columns denote: 
(1) HD number. (2) Identifier from \citet{RM09}. (3) Right ascension (J2000; GAIA DR2). (4) Declination (J2000; GAIA DR2). 
(5) Spectral classification from \citet{RM09}. (6) Mean $V$ magnitude. (7) Mean $V$ magnitude error.
(8) Parallax (Hipparcos). (9) Parallax error.
(10) Parallax (Gaia DR2). (11) Parallax error.
(12) Parallax (Gaia EDR3). (13) Parallax error.
(14) Absorption in the $V$ band. 
(15) Mean effective temperature. (16) Mean effective temperature error.
(17) Luminosity. (18) Luminosity error.
(19) Mass. (20) Evolutionary Phase.}  
\label{table_master1}
\begin{center}
\begin{adjustbox}{max width=\textwidth,angle=90}
\begin{tabular}{lllcclcccccccccccccc}
\hline
\hline
(1) & (2) & (3) & (4) & (5) & (6) & (7) & (8) & (9) & (10) & (11) & (12) & (13) & (14) & (15) & (16) 
& (17) & (18) & (19) & (20)\\
HD	&	ID\_RM09	&	RA(J2000) 	&	 Dec(J2000)    	&	SpT\_RM09	& $V$\,mag	&	e\_$V$\,mag	&
$\pi$ (Hip)	&	e\_$\pi$ (Hip) & $\pi$ (DR2)	&	e\_$\pi$ (DR2) & $\pi$ (EDR3)	&	e\_$\pi$ (EDR3) & $A_V$	& $\log T_\mathrm{eff}$ &
e\_$\log T_\mathrm{eff}$ & $\log L/L_\odot$ & e\_$\log L/L_\odot$ & Mass & Phase \\
\hline
1280 & 270 & 00 17 05.496 & +38 40 53.91 & A2 Si Sr & 4.607 & 0.005 & 10.56 & 0.96 & 15.685 & 0.383 & 18.875 & 0.425 & 0.000 & 3.944 & 0.014 & 1.67 & 0.02 & 2.16 & 5.65 \\
2054 & 490 & 00 25 06.402 & +53 02 48.36 & B9 Si & 5.760 & 0.010 & 6.32 & 0.33 & 5.839 & 0.138 & 5.059 & 0.142 & 0.088 & 4.053 & 0.008 & 2.26 & 0.02 & 3.43 & 5.84 \\
3980 & 1090 & 00 41 46.399 & $-$56 30 04.83 & A7 Sr Eu Cr & 5.698 & 0.003 & 14.92 & 0.35 & 14.613 & 0.090 & 14.770 & 0.041 & 0.000 & 3.914 & 0.005 & 1.28 & 0.01 & 1.91 & 5.57 \\
4853 & 1280 & 00 54 53.111 & +83 42 26.60 & A3 Si & 5.607 & 0.015 & 11.40 & 0.61 & 12.158 & 0.110 & 12.826 & 0.076 & 0.000 & 3.932 & 0.009 & 1.48 & 0.01 & 2.08 & 5.63 \\
5737 & 1490 & 00 58 36.358 & $-$29 21 26.63 & B6 He wk. & 4.310 & 0.010 & 4.20 & 0.18 & 4.707 & 0.423 & 4.622 & 0.191 & 0.015 & 4.141 & 0.005 & 3.21 & 0.08 & 5.40 & 6.48 \\
6397 & 1637 & 01 05 05.356 & +14 56 45.68 & F3 Sr & 5.700 & 0.010 & 18.32 & 0.32 & 19.028 & 0.128 & 18.941 & 0.067 & 0.000 & 3.828 & 0.022 & 1.15 & 0.01 & 1.73 & 5.85 \\
10221 & 2550 & 01 42 20.514 & +68 02 34.85 & A0 Si Sr Cr & 5.576 & 0.019 & 8.39 & 0.50 & 7.682 & 0.113 & 7.696 & 0.057 & 0.002 & 4.029 & 0.015 & 2.02 & 0.02 & 2.86 & 5.68 \\
11502 & 2930 & 01 53 31.814 & +19 17 38.13 & A1 Si Cr Sr & 3.880 & 0.000 & 19.88 & 0.96 & 19.983 & 0.346 & 19.627 & 0.153 & 0.000 & 3.997 & 0.015 & 1.81 & 0.02 & 2.57 & 5.66 \\
11753 & 3010 & 01 54 22.018 & $-$42 29 48.89 & A0 Hg Pt & 5.108 & 0.005 & 10.63 & 0.37 & 10.483 & 0.247 & 10.185 & 0.189 & 0.000 & 3.998 & 0.004 & 1.88 & 0.02 & 2.67 & 5.73 \\
12767 & 3260 & 02 04 29.434 & $-$29 17 48.47 & A0 Si & 4.686 & 0.005 & 8.79 & 0.26 & 9.616 & 0.263 & 9.403 & 0.133 & 0.000 & 4.112 & 0.011 & 2.36 & 0.02 & 3.62 & 5.60 \\
14392 & 3620 & 02 20 58.201 & +50 09 05.44 & B9 Si & 5.550 & 0.010 & 8.31 & 0.34 & 7.899 & 0.109 & 8.528 & 0.168 & 0.009 & 4.069 & 0.020 & 2.09 & 0.01 & 3.01 & 5.45 \\
15089 & 3760 & 02 29 03.941 & +67 24 08.46 & A4 Sr & 4.503 & 0.026 & 24.55 & 0.81 & 21.960 & 0.333 & 22.080 & 0.118 & 0.000 & 3.927 & 0.011 & 1.41 & 0.02 & 2.04 & 5.63 \\
18296 & 4560 & 02 57 17.278 & +31 56 03.30 & A0 Si Sr & 5.102 & 0.004 & 10.20 & 0.27 & 9.848 & 0.258 & 9.345 & 0.132 & 0.090 & 4.041 & 0.009 & 2.05 & 0.02 & 2.99 & 5.69 \\
18519 & 4630 & 02 59 12.719 & +21 20 25.09 & A3 Ti & 5.556 & 0.008 & 9.81 & 0.79 & 8.511 & 0.342 & 9.030 & 0.185 & 0.000 & 3.944 & 0.014 & 1.82 & 0.04 & 2.43 & 5.82 \\
19400 & 4850 & 03 02 15.449 & $-$71 54 09.04 & B8 He wk. & 5.520 & 0.007 & 6.34 & 0.20 & 6.536 & 0.099 & 6.500 & 0.062 & 0.000 & 4.132 & 0.003 & 2.41 & 0.01 & 3.77 & 5.51 \\
19832 & 4910 & 03 12 14.248 & +27 15 25.08 & B8 Si & 5.727 & 0.090 & 6.49 & 0.76 & 8.041 & 0.172 & 7.867 & 0.065 & 0.038 & 4.088 & 0.013 & 2.06 & 0.04 & 3.14 & 5.36 \\
21699 & 5410 & 03 32 08.596 & +48 01 24.58 & B8 He wk. Si & 5.465 & 0.004 & 5.39 & 0.31 & 5.912 & 0.151 & 5.638 & 0.154 & 0.146 & 4.178 & 0.009 & 2.70 & 0.02 & 4.52 & 5.59 \\
22470 & 5710 & 03 36 17.393 & $-$17 28 01.55 & B9 Si & 5.222 & 0.004 & 6.70 & 0.51 & 7.646 & 0.329 & 8.501 & 0.468 & 0.000 & 4.102 & 0.011 & 2.32 & 0.04 & 3.40 & 5.51 \\
22920 & 5840 & 03 40 38.335 & $-$05 12 38.57 & B8 Si & 5.529 & 0.014 & 6.57 & 0.47 & 5.352 & 0.169 & 5.240 & 0.092 & 0.031 & 4.131 & 0.009 & 2.59 & 0.03 & 4.05 & 5.72 \\
23408 & 6000 & 03 45 49.602 & +24 22 03.72 & B7 He wk. Mn & 3.869 & 0.003 & 8.51 & 0.28 & 9.478 & 0.683 & 7.671 & 0.310 & 0.156 & 4.104 & 0.006 & 2.75 & 0.06 & 4.57 & 6.11 \\
23850 & 6100 & 03 49 09.744 & +24 03 12.15 & B8 He wk. & 3.621 & 0.003 & 8.53 & 0.39 & 8.429 & 0.556 & 8.118 & 0.479 & 0.107 & 4.085 & 0.008 & 2.88 & 0.06 & 4.40 & 6.81 \\
25267 & 6440 & 03 59 55.478 & $-$24 00 58.79 & A0 Si & 4.645 & 0.015 & 9.96 & 0.22 & 10.665 & 0.211 & 9.974 & 0.100 & 0.000 & 4.076 & 0.010 & 2.20 & 0.02 & 3.32 & 5.67 \\
25823 & 6560 & 04 06 36.408 & +27 35 59.51 & B9 Sr Si & 5.194 & 0.005 & 7.76 & 0.36 & 8.205 & 0.169 & 8.067 & 0.106 & 0.000 & 4.102 & 0.010 & 2.27 & 0.02 & 3.44 & 5.54 \\
26961 & 6880 & 04 18 14.605 & +50 17 43.81 & A2 Si & 4.592 & 0.003 & 10.40 & 0.35 & 9.558 & 0.763 & 8.547 & 0.234 & 0.000 & 3.961 & 0.011 & 2.12 & 0.07 & 3.00 & 6.24 \\
27309 & 7010 & 04 19 36.695 & +21 46 24.68 & A0 Si Cr & 5.380 & 0.010 & 10.00 & 0.28 & 11.142 & 0.204 & 11.402 & 0.095 & 0.000 & 4.075 & 0.011 & 1.87 & 0.02 & 2.85 & 5.10 \\
28319 & 7240 & 04 28 39.716 & +15 52 15.12 & A8 Sr & 3.398 & 0.004 & 21.69 & 0.46 & 20.835 & 0.373 & 21.813 & 0.374 & 0.016 & 3.902 & 0.012 & 1.90 & 0.02 & 2.40 & 7.07 \\
28843 & 7380 & 04 32 37.560 & $-$03 12 34.39 & B9 He wk. Si & 5.799 & 0.020 & 6.86 & 0.35 & 5.903 & 0.114 & 5.896 & 0.061 & 0.053 & 4.162 & 0.006 & 2.49 & 0.02 & 4.05 & 5.36 \\
29009 & 7420 & 04 33 54.733 & $-$06 44 20.10 & B9 Si & 5.713 & 0.004 & 3.94 & 0.52 & 3.722 & 0.127 & 4.181 & 0.076 & 0.000 & 4.091 & 0.007 & 2.73 & 0.03 & 3.94 & 5.86 \\
29305 & 7490 & 04 33 59.785 & $-$55 02 42.10 & A0 Si & 3.262 & 0.004 & 19.34 & 0.31 &  &  &  &  & 0.034 & 4.064 & 0.002 & 2.22 & 0.01 & 3.23 & 5.70 \\
30612 & 7940 & 04 43 03.948 & $-$70 55 51.79 & B9 Si & 5.523 & 0.006 & 6.64 & 0.23 & 6.841 & 0.075 & 6.970 & 0.063 & 0.006 & 4.094 & 0.008 & 2.28 & 0.01 & 3.39 & 5.58 \\
32549 & 8280 & 05 04 34.140 & +15 24 15.01 & B9 Si Cr & 4.680 & 0.010 & 8.93 & 0.24 & 9.343 & 0.394 & 7.807 & 0.140 & 0.000 & 3.996 & 0.013 & 2.15 & 0.04 & 3.14 & 6.25 \\
32650 & 8350 & 05 12 22.433 & +73 56 47.93 & B9 Si & 5.447 & 0.005 & 8.63 & 0.47 & 10.529 & 0.228 & 8.935 & 0.230 & 0.000 & 4.062 & 0.011 & 1.86 & 0.02 & 2.97 & 5.48 \\
34968 & 8933 & 05 20 26.906 & $-$21 14 22.91 & B9 He var. & 4.703 & 0.005 & 7.84 & 0.51 & 8.719 & 0.166 & 7.822 & 0.112 & 0.000 & 4.005 & 0.010 & 2.21 & 0.02 & 3.08 & 6.56 \\
35039 & 8953 & 05 21 45.751 & $-$00 22 56.87 & B2 He & 4.729 & 0.012 & 3.51 & 0.49 & 2.867 & 0.351 & 2.859 & 0.158 & 0.128 & 4.277 & 0.012 & 3.88 & 0.11 & 8.00 & 7.08 \\
35497 & 9110 & 05 26 17.514 & +28 36 27.12 & B8 Cr Mn & 1.650 & 0.008 & 24.36 & 0.34 &  &  &  &  & 0.000 & 4.098 & 0.014 & 2.73 & 0.01 & 4.17 & 5.90 \\
36485 & 9440 & 05 32 00.413 & $-$00 17 04.19 & B2 He & 6.847 & 0.022 &  &  & 2.573 & 0.077 & 2.625 & 0.054 & 0.135 & 4.246 & 0.009 & 3.05 & 0.03 & 5.57 & 5.46 \\
38104 & 10240 & 05 45 54.039 & +49 49 34.56 & A1 Cr Eu & 5.461 & 0.010 & 7.89 & 0.84 & 5.560 & 0.199 & 8.188 & 0.264 & 0.000 & 3.959 & 0.010 & 2.24 & 0.03 & 2.60 & 5.85 \\
39317 & 10560 & 05 52 22.300 & +14 10 18.45 & B9 Si Eu Cr & 5.550 & 0.010 & 7.69 & 0.29 & 7.143 & 0.176 & 6.855 & 0.096 & 0.000 & 3.991 & 0.011 & 2.02 & 0.02 & 2.85 & 5.85 \\
40312 & 10750 & 05 59 43.245 & +37 12 45.28 & A0 Si & 2.645 & 0.029 & 19.70 & 0.16 & 17.183 & 0.411 &  &  & 0.000 & 4.010 & 0.013 & 2.45 & 0.02 & 3.33 & 6.91 \\
40394 & 10790 & 06 00 58.568 & +47 54 06.94 & A0 Si Fe & 5.732 & 0.024 & 2.82 & 0.47 & 3.588 & 0.127 & 3.155 & 0.082 & 0.185 & 4.003 & 0.012 & 2.64 & 0.03 & 3.86 & 7.23 \\
42509 & 11320 & 06 12 01.340 & +19 47 26.00 & B9 Si & 5.770 & 0.010 & 3.92 & 0.47 & 3.030 & 0.189 & 3.086 & 0.209 & 0.000 & 4.000 & 0.010 & 2.69 & 0.05 & 3.66 & 7.19 \\
47144 & 12660 & 06 35 24.175 & $-$36 46 47.54 & B9 Si & 5.590 & 0.000 & 6.40 & 0.35 & 5.924 & 0.074 & 6.264 & 0.062 & 0.000 & 4.102 & 0.016 & 2.40 & 0.01 & 3.53 & 5.62 \\
47152 & 12670 & 06 38 23.003 & +28 59 03.81 & A0 Eu Cr Hg & 5.759 & 0.043 & 9.41 & 0.62 & 8.514 & 0.137 & 8.274 & 0.157 & 0.129 & 3.992 & 0.010 & 1.84 & 0.02 & 2.61 & 5.72 \\
54118 & 14860 & 07 04 18.331 & $-$56 44 59.02 & A0 Si & 5.151 & 0.002 & 10.84 & 0.43 & 11.683 & 0.182 & 10.363 & 0.175 & 0.000 & 4.009 & 0.020 & 1.79 & 0.01 & 2.67 & 5.66 \\
55719 & 15140 & 07 12 15.790 & $-$40 29 55.66 & A3 Sr Cr Eu & 5.302 & 0.004 & 7.93 & 0.38 & 7.496 & 0.152 & 7.564 & 0.095 & 0.000 & 3.945 & 0.013 & 2.03 & 0.02 & 2.60 & 7.05 \\
56022 & 15250 & 07 13 13.359 & $-$45 10 57.84 & A0 Si Cr Sr & 4.857 & 0.074 & 17.69 & 0.14 & 17.455 & 0.131 & 17.362 & 0.077 & 0.000 & 3.985 & 0.006 & 1.52 & 0.03 & 2.27 & 5.40 \\
56455 & 15400 & 07 14 46.009 & $-$46 50 58.94 & A0 Si & 5.708 & 0.007 & 7.02 & 0.29 & 7.051 & 0.097 & 7.369 & 0.069 & 0.096 & 4.100 & 0.005 & 2.23 & 0.01 & 3.34 & 5.44 \\
59256 & 16180 & 07 27 59.162 & $-$29 09 21.21 & B9 Si & 5.540 & 0.010 & 3.54 & 0.39 & 4.430 & 0.146 & 4.419 & 0.069 & 0.000 & 3.993 & 0.013 & 2.45 & 0.03 & 3.40 & 6.46 \\
59635 & 16280 & 07 29 05.694 & $-$38 48 43.41 & B5 Si & 5.396 & 0.003 & 6.06 & 0.22 & 5.239 & 0.116 & 5.084 & 0.087 & 0.000 & 4.135 & 0.014 & 2.66 & 0.02 & 4.20 & 5.77 \\
61110 & 16715 & 07 39 09.930 & +34 35 03.69 & F2 Sr & 4.902 & 0.012 & 19.61 & 0.30 & 19.742 & 0.215 & 19.334 & 0.126 & 0.000 & 3.815 & 0.007 & 1.48 & 0.01 & 1.95 & 7.11 \\
61641 & 16910 & 07 38 43.893 & $-$36 29 48.56 & B3 He wk. & 5.786 & 0.008 & 3.27 & 0.26 & 3.349 & 0.109 & 2.789 & 0.111 & 0.124 & 4.238 & 0.009 & 3.22 & 0.03 & 6.33 & 5.83 \\
64486 & 17690 & 08 04 47.051 & +79 28 46.89 & A0 Si & 5.407 & 0.015 & 10.10 & 0.24 & 9.912 & 0.200 & 9.570 & 0.088 & 0.015 & 4.001 & 0.004 & 1.82 & 0.02 & 2.60 & 5.67 \\
64740 & 17750 & 07 53 03.645 & $-$49 36 46.95 & B2 He & 4.626 & 0.006 & 4.30 & 0.15 & 4.669 & 0.274 & 4.085 & 0.108 & 0.037 & 4.350 & 0.014 & 3.65 & 0.05 & 8.49 & 5.41 \\
66624 & 18410 & 08 02 44.781 & $-$41 18 35.43 & B9 Si & 5.491 & 0.039 & 6.80 & 0.24 & 7.000 & 0.101 & 6.762 & 0.090 & 0.000 & 4.089 & 0.018 & 2.26 & 0.02 & 3.40 & 5.64 \\
68351 & 18900 & 08 13 08.872 & +29 39 23.56 & A0 Si Cr & 5.626 & 0.004 & 4.77 & 0.40 & 3.328 & 0.246 & 4.672 & 0.224 & 0.000 & 4.002 & 0.013 & 2.67 & 0.06 & 3.18 & 7.06 \\
68601 & 19020 & 08 11 25.901 & $-$42 59 14.22 & A3 Sr & 4.748 & 0.004 & 0.78 & 0.14 & 0.450 & 0.228 & 0.727 & 0.109 & 0.252 & 3.920 & 0.028 & 4.78 & 0.44 & 11.40 & 7.85 \\
71046 & 19636 & 08 19 48.959 & $-$71 30 53.53 & B9 He wk & 5.363 & 0.005 & 7.50 & 0.35 & 7.996 & 0.154 & 7.398 & 0.082 & 0.086 & 4.074 & 0.010 & 2.20 & 0.02 & 3.31 & 5.68 \\
72968 & 20240 & 08 35 28.202 & $-$07 58 56.41 & A2 Sr Cr Eu & 5.715 & 0.005 & 10.77 & 0.36 & 9.367 & 0.107 & 9.628 & 0.048 & 0.000 & 3.978 & 0.015 & 1.70 & 0.01 & 2.39 & 5.63 \\
73340 & 20360 & 08 35 52.032 & $-$50 58 10.81 & B9 Si & 5.798 & 0.009 & 7.31 & 0.25 & 6.705 & 0.115 & 6.702 & 0.066 & 0.028 & 4.126 & 0.019 & 2.28 & 0.02 & 3.57 & 5.36 \\
74067 & 20630 & 08 40 19.180 & $-$40 15 50.04 & A0 Cr Si & 5.193 & 0.012 & 11.68 & 0.50 & 13.001 & 0.138 & 13.197 & 0.228 & 0.000 & 3.993 & 0.016 & 1.65 & 0.01 & 2.39 & 5.49 \\
74521 & 20790 & 08 44 45.039 & +10 04 54.03 & A1 Si Eu Cr & 5.642 & 0.008 & 7.71 & 0.34 & 6.327 & 0.103 & 6.506 & 0.075 & 0.000 & 4.029 & 0.016 & 2.16 & 0.02 & 3.01 & 5.77 \\
74560 & 20840 & 08 42 25.385 & $-$53 06 50.19 & B3 Mg Si & 4.822 & 0.004 & 6.73 & 0.17 & 7.114 & 0.230 & 6.651 & 0.093 & 0.015 & 4.163 & 0.018 & 2.71 & 0.03 & 4.47 & 5.70 \\
77350 & 21860 & 09 02 44.268 & +24 27 10.36 & B9 Sr Cr Hg & 5.445 & 0.017 & 8.31 & 0.35 & 7.737 & 0.165 & 7.295 & 0.148 & 0.000 & 3.989 & 0.006 & 1.99 & 0.02 & 2.83 & 5.85 \\
77653 & 21970 & 09 01 44.584 & $-$52 11 19.42 & B9 Si & 5.225 & 0.005 & 8.85 & 0.42 &  &  &  &  & 0.000 & 4.091 & 0.015 & 2.17 & 0.04 & 3.26 & 5.48 \\
78556 & 22250 & 09 08 42.180 & $-$08 35 22.37 & B9 Si & 5.595 & 0.009 & 5.31 & 0.61 & 3.114 & 0.357 & 4.872 & 0.126 & 0.028 & 4.008 & 0.007 & 2.77 & 0.10 & 3.14 & 7.04 \\
78702 & 22310 & 09 09 04.216 & $-$18 19 42.86 & A1 Si & 5.724 & 0.004 & 11.95 & 0.39 & 9.550 & 0.151 & 9.810 & 0.084 & 0.048 & 3.967 & 0.007 & 1.69 & 0.01 & 2.35 & 5.67 \\
79158 & 22480 & 09 13 48.224 & +43 13 04.27 & B9 He wk. & 5.320 & 0.000 & 5.61 & 0.31 & 5.725 & 0.180 & 5.269 & 0.158 & 0.000 & 4.114 & 0.008 & 2.56 & 0.03 & 4.02 & 5.80 \\
79416 & 22530 & 09 12 30.465 & $-$43 36 47.67 & B8 Si & 5.567 & 0.007 & 4.71 & 0.46 & 5.308 & 0.060 & 5.272 & 0.045 & 0.052 & 4.099 & 0.012 & 2.51 & 0.01 & 3.77 & 5.78 \\
81188 & 23050 & 09 22 06.817 & $-$55 00 38.67 & B3 He wk. & 2.490 & 0.012 & 5.70 & 0.30 &  &  &  &  & 0.097 & 4.260 & 0.010 & 4.12 & 0.05 & 9.44 & 7.27 \\
82984 & 23640 & 09 33 44.508 & $-$49 00 18.71 & B3 He wk. & 5.115 & 0.009 & 3.93 & 0.49 & 3.790 & 0.171 & 3.794 & 0.117 & 0.092 & 4.177 & 0.003 & 3.20 & 0.04 & 5.52 & 5.91 \\
90264 & 25960 & 10 22 58.126 & $-$66 54 05.31 & B8 He wk. & 4.970 & 0.010 & 8.12 & 0.18 & 8.156 & 0.199 & 7.755 & 0.099 & 0.023 & 4.154 & 0.006 & 2.51 & 0.02 & 4.08 & 5.52 \\
90972 & 26150 & 10 29 35.377 & $-$30 36 25.46 & B9 Si & 5.552 & 0.004 & 7.51 & 0.48 & 7.187 & 0.220 & 7.008 & 0.157 & 0.061 & 4.015 & 0.008 & 2.08 & 0.03 & 2.95 & 5.80 \\
92664 & 26750 & 10 40 11.427 & $-$65 06 00.68 & B9 Si & 5.501 & 0.003 & 6.23 & 0.24 & 6.494 & 0.115 & 6.542 & 0.066 & 0.000 & 4.138 & 0.006 & 2.44 & 0.02 & 3.83 & 5.49 \\
92728 & 26760 & 10 43 43.344 & +57 11 57.21 & A0 Si & 5.787 & 0.002 & 8.82 & 0.46 & 7.915 & 0.103 & 8.105 & 0.055 & 0.000 & 3.987 & 0.010 & 1.83 & 0.01 & 2.55 & 5.70 \\
93030 & 26850 & 10 42 57.391 & $-$64 23 40.07 & B0 Si N P & 2.730 & 0.002 & 7.16 & 0.21 &  &  &  &  & 0.098 & 4.393 & 0.018 & 4.16 & 0.03 & 11.01 & 5.69 \\
94334 & 27230 & 10 53 58.724 & +43 11 23.88 & A1 Si & 4.693 & 0.028 & 13.24 & 0.50 & 14.073 & 0.459 & 13.291 & 0.155 & 0.000 & 3.982 & 0.004 & 1.77 & 0.03 & 2.54 & 5.72 \\
95418 & 27493 & 11 01 50.467 & +56 22 56.63 & A0$-$ Ba Y & 2.366 & 0.009 & 40.90 & 0.16 & 33.798 & 1.110 & 38.603 & 1.129 & 0.026 & 3.970 & 0.006 & 1.93 & 0.03 & 2.53 & 5.77 \\
96097 & 27670 & 11 05 01.031 & +07 20 09.67 & F3 Sr & 4.628 & 0.006 & 34.49 & 0.20 &  &  &  &  & 0.000 & 3.854 & 0.015 & 1.01 & 0.01 & 1.64 & 5.64 \\
96616 & 27860 & 11 07 16.678 & $-$42 38 19.48 & A3 Sr & 5.142 & 0.004 & 11.89 & 0.39 & 11.285 & 0.151 & 11.255 & 0.099 & 0.001 & 3.964 & 0.004 & 1.76 & 0.01 & 2.45 & 5.75 \\
97633 & 28150 & 11 14 14.425 & +15 25 46.66 & A2 Sr Eu & 3.338 & 0.023 & 19.76 & 0.17 & 18.495 & 0.451 &  &  & 0.000 & 3.949 & 0.027 & 2.03 & 0.02 & 2.60 & 7.03 \\
98664 & 28450 & 11 21 08.216 & +06 01 45.82 & B9 Si & 4.043 & 0.008 & 14.82 & 0.24 & 16.489 & 0.858 & 13.555 & 0.281 & 0.000 & 3.996 & 0.008 & 1.91 & 0.05 & 2.89 & 5.84 \\
101189 & 29150 & 11 38 07.282 & $-$61 49 35.40 & A0 Cr Y Hg & 5.137 & 0.002 & 11.44 & 0.42 & 9.790 & 0.332 & 10.383 & 0.141 & 0.092 & 4.013 & 0.014 & 1.99 & 0.03 & 2.74 & 5.69 \\
\hline
\end{tabular}  
\end{adjustbox}                                                                                                                                          
\end{center}                                                                                                                                             
\end{table*}

\setcounter{table}{0}  
\begin{table*}
\caption{Essential data for our sample stars, sorted by increasing right ascension. The columns denote: 
(1) HD number. (2) Identifier from \citet{RM09}. (3) Right ascension (J2000; GAIA DR2). (4) Declination (J2000; GAIA DR2). 
(5) Spectral classification from \citet{RM09}. (6) Mean $V$ magnitude. (7) Mean $V$ magnitude error.
(8) Parallax (Hipparcos). (9) Parallax error.
(10) Parallax (Gaia DR2). (11) Parallax error.
(12) Parallax (Gaia EDR3). (13) Parallax error.
(14) Absorption in the $V$ band. 
(15) Mean effective temperature. (16) Mean effective temperature error.
(17) Luminosity. (18) Luminosity error.
(19) Mass. (20) Evolutionary Phase.}  
\label{table_master2}
\scriptsize
\begin{center}
\begin{adjustbox}{max width=\textwidth,angle=90}
\begin{tabular}{lllcclcccccccccccccc}
\hline
\hline
(1) & (2) & (3) & (4) & (5) & (6) & (7) & (8) & (9) & (10) & (11) & (12) & (13) & (14) & (15) & (16) 
& (17) & (18) & (19) & (20)\\
HD	&	ID\_RM09	&	RA(J2000) 	&	 Dec(J2000)    	&	SpT\_RM09	& $V$\,mag	&	e\_$V$\,mag	&
$\pi$ (Hip)	&	e\_$\pi$ (Hip) & $\pi$ (DR2)	&	e\_$\pi$ (DR2) & $\pi$ (EDR3)	&	e\_$\pi$ (EDR3) & $A_V$	& $\log T_\mathrm{eff}$ &
e\_$\log T_\mathrm{eff}$ & $\log L/L_\odot$ & e\_$\log L/L_\odot$ & Mass & Phase \\
\hline
103192 & 29760 & 11 52 54.543 & $-$33 54 29.35 & B9 Si & 4.278 & 0.004 & 10.53 & 0.60 &  &  &  &  & 0.000 & 4.047 & 0.008 & 2.30 & 0.05 & 3.29 & 5.81 \\
106661 & 30870 & 12 16 00.181 & +14 53 56.61 & A2 Si & 5.097 & 0.009 & 16.42 & 0.20 & 15.387 & 0.174 & 15.331 & 0.107 & 0.023 & 3.944 & 0.015 & 1.50 & 0.01 & 2.15 & 5.63 \\
107696 & 31220 & 12 22 49.422 & $-$57 40 33.95 & B8 Cr & 5.380 & 0.008 & 9.02 & 0.29 & 9.692 & 0.157 & 9.609 & 0.072 & 0.023 & 4.079 & 0.015 & 2.01 & 0.01 & 3.04 & 5.35 \\
108283 & 31440 & 12 26 24.064 & +27 16 05.69 & A9 Sr & 4.910 & 0.002 & 11.82 & 0.24 & 12.256 & 0.233 & 11.283 & 0.109 & 0.155 & 3.874 & 0.026 & 1.83 & 0.02 & 2.42 & 7.15 \\
108662 & 31550 & 12 28 54.705 & +25 54 46.33 & A0 Sr Cr Eu & 5.242 & 0.004 & 13.72 & 0.25 & 13.538 & 0.225 & 12.999 & 0.372 & 0.000 & 3.992 & 0.004 & 1.59 & 0.02 & 2.38 & 5.49 \\
108945 & 31610 & 12 31 00.569 & +24 34 01.84 & A3 Sr Cr & 5.433 & 0.003 & 12.09 & 0.27 & 11.997 & 0.139 & 12.079 & 0.121 & 0.000 & 3.951 & 0.003 & 1.57 & 0.01 & 2.23 & 5.66 \\
109026 & 31640 & 12 32 28.003 & $-$72 07 58.95 & B5 He wk. & 3.860 & 0.008 & 10.04 & 0.13 & 9.796 & 0.383 & 8.602 & 0.280 & 0.002 & 4.184 & 0.005 & 2.86 & 0.03 & 5.01 & 5.78 \\
112185 & 32580 & 12 54 01.706 & +55 57 34.98 & A1 Cr Eu Mn & 1.763 & 0.012 & 39.51 & 0.20 & 41.216 & 1.839 &  &  & 0.000 & 3.967 & 0.008 & 1.99 & 0.04 & 2.72 & 5.87 \\
115735 & 33470 & 13 18 14.514 & +49 40 55.38 & B9 He wk. & 5.139 & 0.010 & 11.87 & 0.23 & 11.756 & 0.134 & 11.698 & 0.090 & 0.000 & 4.021 & 0.022 & 1.81 & 0.01 & 2.62 & 5.52 \\
116458 & 33590 & 13 25 50.288 & $-$70 37 38.05 & A0 Eu Cr & 5.663 & 0.004 & 7.35 & 0.30 & 7.759 & 0.181 & 7.992 & 0.117 & 0.000 & 4.004 & 0.021 & 1.93 & 0.02 & 2.68 & 5.70 \\
116656 & 33650 & 13 23 55.632 & +54 55 29.25 & A2 Sr Si & 2.053 & 0.009 & 38.01 & 1.71 &  &  & 40.213 & 0.462 & 0.000 & 3.951 & 0.010 & 1.92 & 0.04 & 2.60 & 6.04 \\
118022 & 34020 & 13 34 07.949 & +03 39 32.43 & A2 Cr Eu Sr & 4.929 & 0.011 & 17.65 & 0.20 & 17.162 & 0.229 & 16.494 & 0.122 & 0.068 & 3.976 & 0.011 & 1.52 & 0.01 & 2.27 & 5.50 \\
120198 & 34660 & 13 46 35.679 & +54 25 57.63 & A0 Eu Cr Sr & 5.685 & 0.021 & 11.23 & 0.23 & 10.619 & 0.069 & 10.742 & 0.042 & 0.000 & 4.003 & 0.018 & 1.64 & 0.01 & 2.41 & 5.41 \\
120640 & 34740 & 13 51 47.221 & $-$46 53 55.14 & B3 He & 5.769 & 0.012 & 2.05 & 0.36 & 2.879 & 0.102 & 3.069 & 0.092 & 0.106 & 4.264 & 0.011 & 3.41 & 0.03 & 6.48 & 5.71 \\
120709 & 34750 & 13 51 49.605 & $-$32 59 38.73 & B5 He wk. P & 4.565 & 0.010 & 9.49 & 0.89 & 11.098 & 0.427 & 9.910 & 0.165 & 0.101 & 4.216 & 0.021 & 2.60 & 0.03 & 4.75 & 5.19 \\
123255 & 35336 & 14 06 42.833 & $-$09 18 48.63 & F1 Cr & 5.464 & 0.011 & 17.23 & 0.37 & 17.504 & 0.148 & 17.563 & 0.099 & 0.049 & 3.850 & 0.019 & 1.29 & 0.01 & 1.86 & 5.84 \\
124224 & 35560 & 14 12 15.805 & +02 24 33.81 & B9 Si & 5.011 & 0.011 & 12.63 & 0.21 & 13.937 & 0.260 & 13.228 & 0.129 & 0.000 & 4.078 & 0.008 & 1.83 & 0.02 & 2.89 & 5.12 \\
124425 & 35610 & 14 13 40.770 & $-$00 50 43.49 & F7 Mg Ca Sr & 5.907 & 0.017 & 16.84 & 0.37 & 18.012 & 0.094 & 18.127 & 0.049 & 0.000 & 3.802 & 0.002 & 1.20 & 0.01 & 1.73 & 6.36 \\
125823 & 35910 & 14 23 02.237 & $-$39 30 42.60 & B5 He wk. & 4.379 & 0.044 & 7.13 & 0.16 & 10.130 & 0.439 & 8.329 & 0.180 & 0.058 & 4.253 & 0.011 & 2.83 & 0.04 & 5.60 & 5.35 \\
128898 & 36710 & 14 42 30.399 & $-$64 58 30.54 & A9 Sr Eu & 3.177 & 0.010 & 60.35 & 0.14 & 62.944 & 0.429 & 60.993 & 0.226 & 0.077 & 3.898 & 0.006 & 1.05 & 0.01 & 1.75 & 5.45 \\
128974 & 36730 & 14 41 01.391 & $-$36 08 05.40 & A0 Si & 5.664 & 0.005 & 5.27 & 0.31 & 4.681 & 0.119 & 4.767 & 0.069 & 0.077 & 4.056 & 0.005 & 2.50 & 0.02 & 3.59 & 5.88 \\
130109 & 37075 & 14 46 14.930 & +01 53 34.44 & A0 Si & 3.735 & 0.012 & 24.25 & 0.18 & 25.138 & 0.371 & 24.281 & 0.227 & 0.086 & 3.974 & 0.006 & 1.67 & 0.01 & 2.40 & 5.67 \\
130158 & 37080 & 14 47 22.559 & $-$25 37 27.23 & B9 Si & 5.611 & 0.000 & 4.39 & 0.34 & 5.682 & 0.174 & 5.982 & 0.074 & 0.092 & 4.019 & 0.007 & 2.28 & 0.03 & 3.14 & 5.86 \\
131120 & 37270 & 14 52 51.080 & $-$37 48 11.38 & B7 He wk. & 5.021 & 0.003 & 6.63 & 0.22 & 6.956 & 0.401 & 6.964 & 0.166 & 0.087 & 4.244 & 0.008 & 2.89 & 0.05 & 5.32 & 5.24 \\
133880 & 38010 & 15 08 12.134 & $-$40 35 02.27 & B9 Si & 5.791 & 0.010 & 9.03 & 0.33 & 9.652 & 0.244 & 9.492 & 0.079 & 0.000 & 4.081 & 0.026 & 1.84 & 0.02 & 2.88 & 5.06 \\
134759 & 38230 & 15 12 13.285 & $-$19 47 30.18 & B9 Si & 4.533 & 0.005 & 8.59 & 0.25 & 7.650 & 0.405 &  &  & 0.021 & 4.058 & 0.013 & 2.50 & 0.05 & 3.65 & 5.89 \\
135382 & 38430 & 15 18 54.601 & $-$68 40 46.24 & A1 Eu & 2.877 & 0.008 & 17.74 & 0.12 & 16.496 & 0.730 & 17.181 & 0.333 & 0.060 & 3.964 & 0.012 & 2.36 & 0.04 & 3.08 & 7.13 \\
136933 & 38890 & 15 24 45.010 & $-$39 42 36.79 & A0 Si & 5.373 & 0.005 & 8.02 & 0.62 & 8.367 & 0.714 & 8.273 & 0.112 & 0.000 & 4.072 & 0.011 & 2.11 & 0.07 & 3.13 & 5.54 \\
137909 & 39200 & 15 27 49.722 & +29 06 20.28 & A9 Sr Eu Cr & 3.678 & 0.013 & 29.17 & 0.76 & 34.867 & 0.989 & 27.926 & 0.970 & 0.056 & 3.887 & 0.012 & 1.36 & 0.03 & 2.14 & 5.87 \\
138764 & 39510 & 15 34 26.509 & $-$09 11 00.29 & B6 Si & 5.166 & 0.009 & 8.17 & 0.58 & 8.729 & 0.227 & 8.489 & 0.145 & 0.094 & 4.098 & 0.020 & 2.26 & 0.02 & 3.43 & 5.58 \\
138769 & 39515 & 15 35 53.240 & $-$44 57 30.06 & B3 He wk & 4.538 & 0.004 & 7.62 & 0.43 & 7.311 & 0.466 & 6.846 & 0.132 & 0.000 & 4.223 & 0.008 & 2.95 & 0.06 & 5.32 & 5.60 \\
140160 & 39840 & 15 41 47.415 & +12 50 51.19 & A1 Sr & 5.322 & 0.017 & 14.84 & 0.41 & 14.551 & 0.133 & 14.722 & 0.075 & 0.000 & 3.957 & 0.004 & 1.46 & 0.01 & 2.13 & 5.50 \\
140728 & 39990 & 15 42 50.772 & +52 21 39.26 & A0 Si Cr & 5.510 & 0.010 & 10.92 & 0.18 & 10.806 & 0.081 & 10.894 & 0.050 & 0.000 & 4.004 & 0.015 & 1.70 & 0.01 & 2.47 & 5.47 \\
142096 & 40340 & 15 53 20.057 & $-$20 10 01.21 & B3 He wk. & 5.031 & 0.006 & 10.54 & 0.91 & 8.565 & 0.317 & 7.388 & 0.290 & 0.461 & 4.214 & 0.008 & 2.77 & 0.03 & 5.06 & 5.55 \\
142990 & 40530 & 15 58 34.871 & $-$24 49 53.19 & B6 He wk. & 5.424 & 0.009 & 5.87 & 0.24 & 6.841 & 0.205 & 7.005 & 0.091 & 0.260 & 4.226 & 0.009 & 2.76 & 0.03 & 4.90 & 5.16 \\
143699 & 40720 & 16 03 24.183 & $-$38 36 09.28 & B6 He wk. & 4.890 & 0.005 & 8.16 & 0.30 & 9.227 & 0.469 & 7.448 & 0.120 & 0.041 & 4.179 & 0.007 & 2.51 & 0.04 & 4.19 & 5.23 \\
148112 & 41850 & 16 25 24.954 & +14 01 59.77 & A0 Cr Eu & 4.570 & 0.014 & 13.04 & 0.64 & 13.515 & 0.358 & 13.103 & 0.166 & 0.000 & 3.977 & 0.009 & 1.84 & 0.02 & 2.59 & 5.78 \\
148330 & 41910 & 16 24 25.334 & +55 12 18.28 & A2 Si Sr & 5.740 & 0.010 & 8.34 & 0.22 & 8.038 & 0.052 & 8.151 & 0.038 & 0.044 & 3.971 & 0.002 & 1.84 & 0.01 & 2.53 & 5.77 \\
148898 & 42070 & 16 32 08.189 & $-$21 27 59.08 & A6 Sr Cr Eu & 4.449 & 0.006 & 19.34 & 0.21 & 19.924 & 0.305 & 19.451 & 0.180 & 0.010 & 3.925 & 0.006 & 1.52 & 0.01 & 2.16 & 5.75 \\
150549 & 42610 & 16 46 40.004 & $-$67 06 34.93 & A0 Si & 5.122 & 0.004 & 4.68 & 0.27 & 5.421 & 0.167 & 4.963 & 0.098 & 0.085 & 4.089 & 0.021 & 2.66 & 0.03 & 4.15 & 5.91 \\
151525 & 42850 & 16 47 46.416 & +05 14 48.24 & B9 Eu Cr & 5.233 & 0.010 & 8.29 & 0.27 & 8.196 & 0.217 & 7.718 & 0.115 & 0.017 & 3.970 & 0.005 & 2.01 & 0.02 & 2.80 & 6.00 \\
152107 & 43050 & 16 49 14.225 & +45 59 00.08 & A3 Sr Cr Eu & 4.820 & 0.011 & 18.10 & 0.34 & 19.136 & 0.199 & 18.759 & 0.096 & 0.000 & 3.961 & 0.025 & 1.42 & 0.01 & 2.14 & 5.47 \\
152127 & 43060 & 16 51 24.921 & +01 12 57.54 & A1 Cr Eu & 5.509 & 0.008 & 8.61 & 0.59 & 6.122 & 0.605 & 8.902 & 0.187 & 0.104 & 3.960 & 0.008 & 2.18 & 0.09 & 2.55 & 5.82 \\
152564 & 43220 & 16 59 33.935 & $-$69 16 05.48 & A0 Si & 5.785 & 0.005 & 4.14 & 0.58 & 4.892 & 0.172 & 5.221 & 0.238 & 0.025 & 4.085 & 0.020 & 2.45 & 0.03 & 3.55 & 5.75 \\
157740 & 44300 & 17 24 31.538 & +16 18 03.56 & A3 Cr Eu Sr & 5.710 & 0.010 & 8.00 & 0.28 & 7.998 & 0.103 & 8.174 & 0.042 & 0.037 & 3.940 & 0.012 & 1.82 & 0.01 & 2.47 & 5.85 \\
166596 & 46900 & 18 13 12.699 & $-$41 20 09.97 & B3 Si & 5.460 & 0.004 & 2.80 & 0.32 & 1.643 & 0.290 & 1.742 & 0.098 & 0.205 & 4.253 & 0.011 & 4.04 & 0.15 & 8.62 & 7.09 \\
168733 & 47280 & 18 22 53.070 & $-$36 40 10.46 & B8 Ti Sr & 5.335 & 0.004 & 5.86 & 0.33 & 5.299 & 0.140 & 5.058 & 0.100 & 0.000 & 4.081 & 0.030 & 2.54 & 0.02 & 3.84 & 5.87 \\
169467 & 47480 & 18 26 58.421 & $-$45 58 06.43 & B4 He & 3.509 & 0.003 & 11.74 & 0.17 & 13.136 & 0.431 & 11.670 & 0.405 & 0.000 & 4.165 & 0.018 & 2.70 & 0.03 & 4.53 & 5.72 \\
170000 & 47620 & 18 20 45.382 & +71 20 15.99 & A0 Si & 4.215 & 0.009 & 10.77 & 0.38 &  &  &  &  & 0.009 & 4.057 & 0.008 & 2.33 & 0.03 & 3.36 & 5.80 \\
172728 & 48420 & 18 37 33.492 & +62 31 35.62 & A0 Hg Y Zr & 5.740 & 0.000 & 8.40 & 0.22 & 7.847 & 0.090 & 7.788 & 0.088 & 0.054 & 4.009 & 0.004 & 1.92 & 0.01 & 2.72 & 5.69 \\
175156 & 49010 & 18 54 43.125 & $-$15 36 10.94 & B5 He var. & 5.092 & 0.010 & 2.33 & 0.24 & 1.513 & 0.352 & 1.767 & 0.126 & 1.042 & 4.174 & 0.011 & 4.38 & 0.20 & 10.45 & 7.53 \\
175362 & 49030 & 18 56 40.503 & $-$37 20 35.47 & B6 He wk. Si & 5.373 & 0.005 & 7.58 & 0.27 & 6.537 & 0.192 & 6.843 & 0.192 & 0.085 & 4.222 & 0.014 & 2.74 & 0.03 & 4.83 & 5.15 \\
177003 & 49360 & 19 00 13.672 & +50 32 00.49 & B3 He var. & 5.373 & 0.004 & 4.86 & 0.18 & 5.581 & 0.155 & 5.106 & 0.092 & 0.042 & 4.247 & 0.015 & 2.93 & 0.02 & 5.55 & 5.42 \\
177756 & 49540 & 19 06 14.940 & $-$04 52 57.18 & B8 Si & 3.432 & 0.008 & 26.37 & 0.64 & 27.060 & 0.508 & 25.641 & 0.421 & 0.000 & 4.036 & 0.003 & 1.79 & 0.02 & 2.71 & 5.44 \\
181615 & 50300 & 19 21 43.634 & $-$15 57 18.20 & B8 He & 4.610 & 0.000 & 1.83 & 0.23 & 1.492 & 0.329 & 2.087 & 0.172 & 0.831 & 4.227 & 0.027 & 4.64 & 0.19 & 11.03 & 7.45 \\
182255 & 50370 & 19 22 50.890 & +26 15 44.76 & B7 He wk. & 5.193 & 0.011 & 8.30 & 0.60 & 9.058 & 0.197 & 8.907 & 0.169 & 0.041 & 4.141 & 0.004 & 2.30 & 0.02 & 3.70 & 5.27 \\
182568 & 50490 & 19 24 07.577 & +29 37 16.93 & B3 He wk. & 4.981 & 0.008 & 3.87 & 0.21 & 3.566 & 0.169 & 3.624 & 0.136 & 0.302 & 4.252 & 0.012 & 3.59 & 0.04 & 6.98 & 5.88 \\
183056 & 50610 & 19 26 09.130 & +36 19 04.42 & B9 Si & 5.155 & 0.005 & 4.61 & 0.22 & 5.836 & 0.137 & 5.777 & 0.101 & 0.000 & 4.080 & 0.012 & 2.53 & 0.02 & 3.74 & 5.84 \\
183806 & 50750 & 19 33 21.622 & $-$45 16 18.37 & A0 Cr Eu Sr & 5.580 & 0.004 & 8.22 & 0.40 & 7.753 & 0.131 & 8.253 & 0.170 & 0.000 & 3.975 & 0.011 & 1.92 & 0.02 & 2.58 & 5.79 \\
185872 & 51260 & 19 39 26.481 & +42 49 05.71 & B9 Si & 5.415 & 0.009 & 5.00 & 0.19 & 5.131 & 0.179 & 6.169 & 0.245 & 0.024 & 4.021 & 0.009 & 2.43 & 0.03 & 3.17 & 5.86 \\
187474 & 51700 & 19 51 50.602 & $-$39 52 27.79 & A0 Eu Cr Si & 5.317 & 0.010 & 10.82 & 0.88 & 13.689 & 0.385 & 10.341 & 0.312 & 0.000 & 4.010 & 0.018 & 1.58 & 0.03 & 2.56 & 5.54 \\
188041 & 51900 & 19 53 18.737 & $-$03 06 52.15 & A6 Sr Cr Eu & 5.640 & 0.000 & 12.48 & 0.36 & 11.725 & 0.101 & 11.867 & 0.062 & 0.000 & 3.916 & 0.015 & 1.49 & 0.01 & 2.10 & 5.74 \\
189178 & 52400 & 19 57 13.879 & +40 22 04.13 & B5 He wk. & 5.455 & 0.015 & 2.94 & 0.19 & 2.857 & 0.103 & 2.831 & 0.070 & 0.277 & 4.162 & 0.001 & 3.35 & 0.03 & 5.80 & 6.45 \\
196178 & 54690 & 20 33 54.844 & +46 41 37.91 & B8 Si & 5.774 & 0.005 & 7.56 & 0.47 & 6.742 & 0.129 & 5.946 & 0.092 & 0.000 & 4.110 & 0.010 & 2.23 & 0.02 & 3.56 & 5.57 \\
196502 & 54780 & 20 31 30.424 & +74 57 16.63 & A2 Sr Cr Eu & 5.192 & 0.004 & 8.24 & 0.47 & 8.082 & 0.201 & 8.579 & 0.297 & 0.000 & 3.937 & 0.018 & 2.00 & 0.02 & 2.60 & 6.32 \\
196524 & 54790 & 20 37 32.919 & +14 35 42.20 & F5 Sr & 3.623 & 0.022 & 32.33 & 0.47 &  &  &  &  & 0.000 & 3.810 & 0.003 & 1.58 & 0.02 & 2.04 & 7.18 \\
198667 & 55300 & 20 52 08.706 & $-$05 30 25.21 & B9 Si & 5.544 & 0.008 & 5.02 & 0.31 & 3.983 & 0.139 & 4.183 & 0.086 & 0.030 & 4.029 & 0.005 & 2.61 & 0.03 & 3.53 & 6.91 \\
201433 & 56170 & 21 08 38.876 & +30 12 20.39 & B9 Si Mg & 5.612 & 0.020 & 8.64 & 0.55 & 8.168 & 0.111 & 8.199 & 0.104 & 0.083 & 4.054 & 0.008 & 2.03 & 0.01 & 2.96 & 5.55 \\
201601 & 56210 & 21 10 20.488 & +10 07 53.71 & A9 Sr Eu & 4.697 & 0.025 & 27.55 & 0.62 & 28.771 & 0.232 & 28.243 & 0.148 & 0.036 & 3.890 & 0.005 & 1.11 & 0.01 & 1.77 & 5.56 \\
201834 & 56290 & 21 10 15.563 & +53 33 47.21 & B9 Si & 5.744 & 0.004 & 7.36 & 0.23 & 6.404 & 0.175 & 7.135 & 0.113 & 0.000 & 4.086 & 0.017 & 2.23 & 0.02 & 3.19 & 5.47 \\
202444 & 56410 & 21 14 47.462 & +38 02 42.94 & F0 Sr & 3.732 & 0.007 & 49.16 & 0.40 & 49.576 & 0.463 &  &  & 0.000 & 3.832 & 0.021 & 1.10 & 0.01 & 1.69 & 5.81 \\
202627 & 56450 & 21 17 56.293 & $-$32 10 21.32 & A1 Si & 4.708 & 0.011 & 17.90 & 0.23 & 19.705 & 0.547 & 19.298 & 0.267 & 0.000 & 3.963 & 0.022 & 1.44 & 0.02 & 2.16 & 5.47 \\
202671 & 56480 & 21 17 57.281 & $-$17 59 06.45 & B7 He wk. Mn & 5.415 & 0.019 & 6.11 & 0.31 & 4.761 & 0.292 & 4.277 & 0.113 & 0.021 & 4.130 & 0.007 & 2.74 & 0.05 & 4.45 & 5.86 \\
203006 & 56550 & 21 20 45.647 & $-$40 48 34.18 & A2 Cr Eu Sr & 4.817 & 0.008 & 16.54 & 1.19 & 18.268 & 0.531 & 20.169 & 0.497 & 0.000 & 3.989 & 0.026 & 1.50 & 0.03 & 2.19 & 5.17 \\
203585 & 56690 & 21 24 24.815 & $-$41 00 24.14 & A0 Si & 5.762 & 0.016 & 8.35 & 0.70 &  &  &  &  & 0.000 & 4.012 & 0.020 & 1.84 & 0.07 & 2.63 & 5.60 \\
205637 & 57290 & 21 37 04.833 & $-$19 27 58.05 & B4 Si & 4.641 & 0.035 & 3.09 & 0.18 & 6.550 & 0.705 & 3.678 & 0.200 & 0.000 & 4.171 & 0.017 & 2.87 & 0.09 & 5.75 & 6.86 \\
206742 & 57530 & 21 44 56.796 & $-$33 01 32.49 & A0 Si & 4.339 & 0.007 & 15.97 & 0.17 & 16.385 & 0.441 & 15.335 & 0.142 & 0.002 & 3.996 & 0.004 & 1.80 & 0.02 & 2.60 & 5.70 \\
209515 & 58290 & 22 02 56.658 & +44 38 59.62 & A0 Cr Si Mg & 5.599 & 0.006 & 6.62 & 0.39 & 6.021 & 0.155 & 6.169 & 0.053 & 0.048 & 3.988 & 0.006 & 2.17 & 0.02 & 2.96 & 5.89 \\
215766 & 59600 & 22 47 42.773 & $-$14 03 23.04 & B9 Si & 5.660 & 0.010 & 10.27 & 0.46 & 9.185 & 0.130 & 9.132 & 0.056 & 0.060 & 4.019 & 0.028 & 1.83 & 0.01 & 2.64 & 5.56 \\
220825 & 60520 & 23 26 55.952 & +01 15 20.15 & A1 Cr Sr Eu & 4.936 & 0.011 & 21.25 & 0.29 & 20.442 & 0.216 & 20.315 & 0.098 & 0.125 & 3.962 & 0.036 & 1.37 & 0.01 & 2.09 & 5.36 \\
221006 & 60580 & 23 29 00.983 & $-$63 06 38.34 & A0 Si & 4.705 & 0.005 & 8.44 & 0.29 & 8.051 & 0.110 & 8.406 & 0.069 & 0.000 & 4.121 & 0.011 & 2.53 & 0.01 & 3.82 & 5.67 \\
221760 & 60770 & 23 35 04.579 & $-$42 36 54.21 & A2 Sr Cr Eu & 4.701 & 0.002 & 13.11 & 0.53 & 12.843 & 0.392 & 12.306 & 0.580 & 0.000 & 3.938 & 0.007 & 1.80 & 0.03 & 2.50 & 5.87 \\
223640 & 61290 & 23 51 21.328 & $-$18 54 32.76 & B9 Si Sr Cr & 5.180 & 0.014 & 10.23 & 0.31 & 9.864 & 0.316 & 9.980 & 0.154 & 0.000 & 4.083 & 0.008 & 2.07 & 0.03 & 3.11 & 5.38 \\
\hline
\end{tabular}  
\end{adjustbox}                                                                                                                                          
\end{center}                                                                                                                                             
\end{table*}

\label{lastpage}
\end{document}